\documentclass[useAMS,usenatbib]{mn2e}

\usepackage{graphicx}% Include figure files
\usepackage{dcolumn}% Align table columns on decimal point
\usepackage{bm}% bold math
\usepackage{amssymb, amsmath}

\usepackage{hyperref}

\providecommand{\adsurl}[1]{\href{#1}{ADS}}

\usepackage{aas_macros}

\newcommand\beq{\begin{equation}}
\newcommand\eeq{\end{equation}}
\newcommand\beqar{\begin{eqnarray}}
\newcommand\eeqar{\end{eqnarray}}

\newcommand{\cla}{\hat{C}_{\ell, {\rm 1}}}
\newcommand{\clb}{\hat{C}_{\ell, {\rm 2}}}
\newcommand{\clc}{\hat{C}_{\ell, {\rm 3}}}
\newcommand{\ina}{I_{\rm 1}}
\newcommand{\inb}{I_{\rm 2}}
\newcommand{\inc}{I_{\rm 3}}
\newcommand{\clo}{\hat{C}_{\ell, {\rm tot}}}
\newcommand{\ino}{I_{\rm tot}}

\newcommand{\clao}{\hat{C}_{\ell_1,1}}
\newcommand{\clat}{\hat{C}_{\ell_2,1}}
\newcommand{\clbo}{\hat{C}_{\ell_1,2}}
\newcommand{\clbt}{\hat{C}_{\ell_2,2}}
\newcommand{\clm}{\hat{C}_{\ell, {\rm m}}}
\newcommand{\cloo}{\hat{C}_{\ell_1,{\rm tot}}}
\newcommand{\clot}{\hat{C}_{\ell_2,{\rm tot}}}

\begin{document}

\title{Novel Techniques for Decomposing Diffuse Backgrounds}
\author[Brandon S. Hensley, Vasiliki Pavlidou, and Jennifer
M. Siegal-Gaskins]{Brandon S. Hensley$^{1}$\thanks{E-mail:
bhensley@princeton.edu}, Vasiliki Pavlidou$^{2,3}$, and Jennifer M. Siegal-Gaskins$^{4,5,6}$\\
$^{1}$Department of Astrophysical Sciences, Princeton
  University, 4 Ivy Lane, Princeton, NJ 08544, USA\\
$^{2}$Max-Planck Institute for Radioastronomy, 53121 Bonn, Germany\\
$^{3}$Current Address: Department of Physics, University of Crete,
71003 Heraklion, Greece\\
$^{4}$Einstein Fellow\\
$^{5}$California Institute of Technology, Pasadena, CA 91125, USA\\
$^{6}$Center for Cosmology and Astro-Particle Physics, Ohio State
University, Columbus, OH 43210, USA
}

\date{\today}% It is always \today, today,

\pagerange{\pageref{firstpage}--\pageref{lastpage}} \pubyear{2012}

\maketitle

\label{firstpage}

\begin{abstract}
The total anisotropy of a diffuse background composed of two or more
sources, such as the \emph{Fermi}-LAT--measured gamma-ray background, is set by the anisotropy of each source population and the
contribution of each population to the total intensity. The total
anisotropy as a function of energy (the anisotropy energy spectrum)
will modulate as the relative contributions of the sources change,
implying that the anisotropy energy spectrum also encodes the
intensity spectrum of each source class.
We develop
techniques, applicable to any such diffuse background, for unraveling the intensity spectrum of each 
component source population given a measurement of the
total intensity spectrum and the total anisotropy energy spectrum,
without introducing \emph{a priori} assumptions about the spectra of the
source classes. We demonstrate the potential of these methods by applying them to example scenarios for the composition of the \emph{Fermi}-LAT gamma-ray background consistent with current data and feasible within 10~years of observation.
\end{abstract}


\section{Introduction}
Diffuse emission, from radio to gamma-ray frequencies, encodes a wealth of information about fundamental physics, cosmology, and a variety of astrophysical systems. Prominent examples include the cosmic microwave background, a snapshot of the very early universe at microwave frequencies~\citep{Komatsu:2010fb}; direct and reprocessed starlight between infrared and ultraviolet wavelengths, a record of the star formation history of the universe~\citep{Primack+Dominguex+Gilmore+Somerville_2011,Kneiske+Dole_2010,Stecker+Baring+Summerlin_2007}; thermal emission from accretion processes in X-rays, which traces the growth of black holes through cosmic time~\citep{Soltan_2007}; and non-thermal emission in gamma rays from, e.g.,  blazars~\citep{Collaboration:2010gqa:srccounts, Abazajian:2010pc, Stecker:2010di}, star-forming galaxies~\citep{Fields:2010bw}, millisecond pulsars~\citep{FaucherGiguere:2009df}, and, possibly, annihilating or decaying dark matter~\citep{Ullio:2002pj,Overduin:2004sz,Bertone:2007aw}. 

Both the intensity spectrum and the degree of anisotropy of the
diffuse signal have been successfully used to uncover valuable
information about the physics and astrophysics of the processes and
sources that are responsible for the diffuse emission in each waveband, the cosmic microwave background
being the most celebrated example of both methods
\citep[e.g.,][]{Mather+Cheng+Eplee+etal_1990, Spergel+Verde+Peiris+etal_2003}. However, when more than one source class contributes to the diffuse emission, complications to such analyses arise. Traditionally, 
determining the individual contributions of source classes in a
multi-population diffuse signal has relied on careful modeling and
subtraction of intensity spectra~\citep{Strong+Moskalenko_Reimer_2004}. This process is hindered by systematic uncertainties in
the theoretical understanding of astrophysical source classes, which
limit our ability to detect a possibly subdominant signal, especially one with unknown
or poorly constrained properties.  

In this work, we
illustrate that a measurement of the anisotropy of a diffuse background at a
fixed angular scale as a function of energy, when combined with the
total intensity
spectrum of the background, can be used to decouple the contributions to the
background of each source population, thereby yielding an intensity spectrum for
each source class. We focus here on applications to the isotropic
diffuse gamma-ray background, as its large range of plausible
compositions allows us to illustrate many of the techniques
presented here.

The isotropic gamma-ray background (IGRB), the diffuse gamma-ray emission at
energies above $\sim 100$~MeV that is isotropic on large angular scales, is one of the most promising observational targets for the discovery of new physics in this decade, such as a signature from dark matter annihilation or decay.  Although the IGRB has been observed since the 1970s \citep{Fichtel77,sreekumar_bertsch_dingus_etal_98}, the Large Area Telescope (LAT) aboard the currently operational Fermi Gamma-Ray Space telescope \citep[\emph{Fermi}][]{Atwood2009} is improving both the energy range and the angular accuracy of these observations.  The LAT also resolves more bright point sources than previous missions due to its increased sensitivity, providing valuable information about gamma-ray source populations via detected members.  As a result, the LAT collaboration has reported a more precise measurement of the IGRB intensity spectrum~\cite{Abdo:2010nz:igrb}, and for the first time has measured the small-scale anisotropy of the IGRB~\cite{Ackermann:2012uf}.

In addition to any possible exotic signal, confirmed astrophysical
gamma-ray sources such as gamma-ray loud active galactic nuclei
(blazars) and star-forming galaxies are guaranteed to contribute
significantly to the IGRB at some energy. \citet{Siegal-Gaskins2009} showed that by
combining the spectral and anisotropy properties of the IGRB, it is
possible to identify the presence of a second, even subdominant,
component, such as a signal from dark matter annihilation or decay, over a dominant, astrophysical contribution. We extend this approach by developing techniques that allow the intensity spectra of the individual components to be reconstructed without requiring a model or prediction for any of the contributions. 

For diffuse backgrounds composed of emission from uncorrelated source populations, we show that under certain conditions,
 if the {\em intensity energy spectrum} (differential photon intensity
 as a function of energy) and the {\em anisotropy energy spectrum}
 (angular power at a fixed multipole as a function of
 energy) of the diffuse background are both measured with sufficient accuracy, the shape of the intensity energy spectrum of each component can be recovered; in some cases the absolute normalisations of the intensity spectra are also recoverable.  Similarly, in some cases the amplitude of the angular power spectra of the individual components can also be determined.  We discuss the conditions under which such decompositions are feasible, and demonstrate 
these novel techniques on plausible scenarios for the IGRB composition.  Although our examples are restricted to the IGRB, the methods
presented here can be applied to any diffuse background at any
wavelength.

In \S\ref{sec:method} we introduce the formalism common to all of the
decomposition techniques.  In \S\ref{sec:examples} we define IGRB
component models and simulated observations used for the example
scenarios.  The details of each decomposition technique are described
in \S\ref{sec:details}; example scenarios illustrating a subset of the
techniques are also presented.  We extend our approach to selected
three-component scenarios in \S\ref{sec:threecomp}.  In
\S\ref{sec:threeastwo}, we explore the systematic errors that arise
when using two-component techniques when in reality the emission comes
from three source classes. Finally, we discuss the potential of these techniques for understanding gamma-ray source populations in \S\ref{sec:disc}.

\section{Two-Component Decomposition: Methods}
\label{sec:method}

The two properties of diffuse emission we will use are the
differential intensity energy spectrum $I(E)$ (photons per  area per
time per solid angle per energy) and the angular power spectrum
$C_\ell$ of a sky map of the intensity.  The angular power spectrum is
defined as $C_\ell = \langle|a_{\ell m}|^2\rangle$, where $a_{\ell m}$
are the coefficients of the expansion of the intensity map in the
basis of spherical harmonics.  We also define the fluctuation angular
power spectrum $\hat{C}_{\ell} \equiv C_\ell / I^{2}$, where $I$ is
the mean intensity of the emission with intensity angular power
spectrum $C_{\ell}$.  Because $\hat{C}_{\ell}$ describes fluctuations
in units of the mean, the fluctuation angular power at a fixed $\ell$
is energy-independent for a signal arising from a single population of
sources with identical observer-frame intensity spectra.  In the following
we assume that each distinct component of the diffuse emission meets
this criterion.

Variation between the source spectra of individual members of a
population can result in  fluctuation angular power which is energy
dependent because the relative contributions of spectrally different
sources within a population change with energy (e.g., harder sources
contribute relatively more flux at high energies than at low
energies).  In addition, for cosmological source populations,
energy-dependent fluctuation angular power can also arise due to
redshifting of sharp features in the source spectra, such as line
emission or abrupt cut-offs~\citep[see, e.g.,][]{zhang_beacom_04, ando_komatsu_06}.  

In practice, if a component of the emission arises from a population
of sources, we assume that the requirement that the single-population $\hat{C}_{\ell}$ is energy-independent is satisfied if the
variation in the intensity spectra of individual members of the
population is sufficiently small that the deviation of the fluctuation
angular power from an energy-independent quantity is at a level
smaller than the uncertainty on the anisotropy measured by a specific
observation.  We comment on the validity of this assumption in the context of the IGRB in \S\ref{sec:examples}.

Our approach exploits the energy independence of the single-component fluctuation angular power, and so it is convenient for us to work with $\hat{C}_{\ell}$.
If we consider a scenario in which the diffuse emission is composed of emission from two spatially uncorrelated components with intensity spectra $I_{1}(E)$ and $I_{2}(E)$ and angular power spectra $C_{\ell,1}$ and $C_{\ell,2}$, then 
the total intensity is simply the sum of the two components,
\begin{equation}\label{one}
\ino(E) = \ina(E) + \inb(E)\,.
\end{equation}
The angular power spectrum of the total signal for uncorrelated components is the sum of the angular power spectra of the components,
\begin{equation}
C_{\ell,{\rm tot}}(E)=C_{\ell,1}(E) + C_{\ell,2}(E).
\end{equation}
Rewritten in terms of the fluctuation angular power, 
\begin{equation}\label{two}
\clo(E) = \left(\frac{\ina(E)}{\ino(E)}\right)^2 \cla
+ \left(\frac{\inb(E)}{\ino(E)}\right)^2 \clb\,.
\end{equation}
This is the fluctuation anisotropy energy spectrum for the case we consider.  In the following we will always use the term ``anisotropy energy spectrum'' to refer to the \emph{fluctuation} angular power of the total emission as a function of energy.

\begin{table*}
\caption{Summary of two-component decomposition techniques. \label{table1}}
%\begin{ruledtabular}
\begin{tabular}%{ccccc}
%\begin{center}
    {  p{2.5cm}  p{4.cm}  p{4.cm}  p{2.0cm}  p{2.0cm} }
    \noalign{\smallskip} Method & Observational Signature & Inferred Properties of
    Components & Intensity \mbox{Normalisation} Recovered? & Fluctuation \mbox{Angular Power} Recovered?\\ \noalign{\smallskip} \hline \hline \noalign{\smallskip}
    Double plateau & Plateaus at both high and low energies observed in anisotropy energy spectrum & One source dominant in anisotropy at low energies, other source dominant at high energies &
    Yes & Yes \\ \noalign{\smallskip} \hline \noalign{\smallskip}
    Low-Anisotropy Plateau & Anisotropy energy spectrum rises
    from (falls to) a low-anisotropy plateau at low (high) energy & Source that is subdominant in intensity is 	much more anisotropic
    than the dominant source  & No & No \\ \noalign{\smallskip} \hline \noalign{\smallskip}
    High-Anisotropy Plateau & Anisotropy energy spectrum falls
    from (rises to) a high-anisotropy plateau at low (high) energy & Source that is subdominant in intensity 		is much less anisotropic than the dominant source & Yes & No \\ \noalign{\smallskip} \hline \noalign{\smallskip}
    Known \mbox{Zero-Anisotropy} Component  & None; requires {\em a priori} knowledge that one of the two 		components is isotropic & One source is completely isotropic & No & No \\ \noalign{\smallskip} \hline \noalign{\smallskip}
    Minimum & Minimum observed in the anisotropy energy spectrum & Both source components have
    comparable intensity and anisotropy such that Eq.~\ref{eq:Min} is satisfied at some
    energy & Yes & Yes \\ \noalign{\smallskip} \hline \noalign{\smallskip}
    Multiple-$\ell$ Measurements & Two distinct anisotropy energy
    spectra can be obtained at two different $\ell$ & $\hat{C}_\ell$ is a function of $\ell$ for at least one source such that two distinct
    anisotropy energy spectra can be obtained at different
    $\ell$ & Yes & Yes \\ \noalign{\smallskip}
    \end{tabular}
%\end{ruledtabular}
%\end{center}
\end{table*}

With sufficient photon statistics, $\ino$ and $\clo$ can be determined at each energy from observations. If there is a way to also determine $\cla$ and $\clb$ from the data, we can solve Eqs. (\ref{one}) and (\ref{two}) for $\ina$ and $\inb$: 
\begin{equation}\label{Ieg}
\ina = \ino \left(\frac{\clb \pm \sqrt{\cla \clo + \clb \clo - \cla \clb}}{\cla + \clb}\right) 
\end{equation}
\begin{equation}\label{Idm}
\inb = \ino \left(\frac{\cla \mp \sqrt{\cla \clo + \clb \clo - \cla
      \clb}}{\cla + \clb}\right) 
~~~.
\end{equation}

If there is an energy $\sim E_0$ around which only one component is expected
to contribute to the total intensity (i.e., an energy range around
$E_0$ where $I_2(E_0)/I_{\rm tot}(E_0)\approx 0$), the anisotropy
energy spectrum will be flat over this energy range. Then from
Eq.~(\ref{two}) we immediately obtain $\cla = \hat{C}_{\ell,{\rm
  tot}}(E_0)$ from the anisotropy of this baseline.  A similar
flat baseline could result if two source classes have the
same spectral shape over an energy range, but such a scenario is
unlikely for the source classes considered here. 
 {\em In each of
the following cases, we will assume either that we can obtain the $\hat{C}_\ell$ of one
of the two source classes in this way, or that one source class is
known to have $\hat{C}_{\ell} \simeq 0$.}

We discuss 
six distinct two-component decomposition techniques below.
Some of them allow us to extract the component intensity spectra, while others only allow us to derive the {\em shapes} 
of one or both intensity spectra up to unknown normalisation constants.  Some of the techniques also yield measurements of the fluctuation angular power spectra of each component source population.
Table \ref{table1} gives a summary of these techniques and
their applicability conditions. We emphasise that in all cases we make the following three assumptions: (1) the diffuse background is composed of emission from uncorrelated source classes, (2) the fluctuation angular power of each individual component is independent of energy, and (3) the fluctuation angular power of one component can be directly measured from the data at some energy or is known to be negligibly small.

\section{Parameters of Example Scenarios}
\label{sec:examples}

\subsection{IGRB Component Models}

To illustrate the decomposition techniques, we apply them to example scenarios that
could be measured by the \emph{Fermi}-LAT within 10~years of
observation time. Each scenario is a two-component IGRB model consistent with current observations of the
measured IGRB intensity energy spectrum~\citep{Abdo:2010nz:igrb} and
the measurement of the IGRB anisotropy energy
spectrum~\citep{Ackermann:2012uf}. In particular, we consider a power
law component with slope and fluctuation anisotropy that describes the
IGRB well at low energies but that may break at high energies, as well
as a Galactic dark
matter annihilation component with one of two benchmark spectra
(annihilation to a $\tau^{+}\tau^{-}$ or $b\bar{b}$ final state). 

The power law component encapsulates  likely contributions from several source
classes such as blazars, but we assume, in accord with the data, that
this emission can be characterised by a single $\hat{C}_\ell$. The
adopted $\hat{C}_\ell$ in this work differs from that derived
in~\citep{Ackermann:2012uf} as we use the intensities reported
by~\citep{Abdo:2010nz:igrb} rather than those used in the anisotropy
analysis, which were subjected to less cleaning. As this cleaning
likely removed only isotropic contamination, we expect that the
angular power $C_P$ derived in~\citep{Ackermann:2012uf} also describes the IGRB
of~\citep{Abdo:2010nz:igrb}, and thus we check for consistency with
$C_P$ only.

Although we do not consider them explicitly in our example scenarios, other known gamma-ray source populations which may contribute significantly to the intensity and/or anisotropy of the IGRB at some energies include star-forming galaxies~\citep{Fields:2010bw, Ando:2009nk, SFlat, SFlacki, SFnachi, SFtoni},  gamma-ray loud radio galaxies~\citep{Inoue:2011bm} and Galactic millisecond pulsars~\citep{FaucherGiguere:2009df, SiegalGaskins:2010mp}.

For source classes relevant for the IGRB, the three assumptions stated above, which are necessary to implement our methods, are very likely to be valid.   While all cosmological populations will generally trace large-scale structure, on the small angular scales considered in the \emph{Fermi} LAT anisotropy analysis~\citep{Ackermann:2012uf} we do not expect strong spatial correlations between, e.g., blazars and extragalactic dark matter~\citep{Ando:2006cr}, and naturally the angular distribution of any Galactic source population is entirely uncorrelated with that of any extragalactic population.  For these source classes, the variation between observer-frame source spectra within the population is likely to be sufficiently small (or zero, in the case of Galactic dark matter annihilation or decay) to induce at most a mild energy dependence in the fluctuation angular power, although we caution that a careful investigation of the expected magnitude of this effect is needed.  Furthermore, most expected cosmological contributors to the IGRB are not expected to exhibit features in their spectra which are sharp enough to lead to significant energy dependence of the fluctuation angular power due to redshifting.  Although in some dark matter models sharp features are present in the photon spectra, in many scenarios the Galactic dark matter signal is expected to dominate over the extragalactic signal in both intensity and anisotropy, so any energy dependence in the fluctuation angular power of the extragalactic dark matter component would likely have a subdominant effect.  Finally, the expected contributors to the IGRB have different spectral shapes, and consequently it is not unlikely that at certain energies all but one component will supply a negligible contribution, in which case the fluctuation angular power of the dominant component could be measured as described in the previous section.

The level of the  blazar contribution to the IGRB is uncertain, with
different calculations spanning a large range of possibilities
\citep[e.g.,][]{inouetotani, Collaboration:2010gqa:srccounts, Abazajian:2010pc, Stecker:2010di, Cuoco:2012yf, Harding:2012gk}. The strongest bounds on the contribution of blazars to the intensity of the IGRB have been obtained by requiring that the adopted model for the blazar population does not exceed the measured IGRB anisotropy~\citep{Cuoco:2012yf, Harding:2012gk}; these bounds limit the blazar IGRB intensity contribution to $\lesssim 20\%$ in the 1--10~GeV band.  The spectral shape of the
blazar contribution is dependent primarily on the distribution of
blazar spectral indices in the gamma-ray range
\citep{PavlidouVenters2007}. Since BL Lac--type blazars and flat-spectrum
radio quasars (FSRQs), the two largest subclasses of blazars, generally have different spectral properties, the shape of the collective intensity spectrum
depends on the relative abundances of these two subclasses in the unresolved blazar
population, which is uncertain due to the difficulties in obtaining BL
Lac redshifts and assessing the prevalence of BL Lacs in the
high-redshift universe \citep{Collaboration:2010gqa:srccounts}. Additional uncertainties enter
through considerations regarding the fraction of blazars with spectral
breaks \cite{tonivaso11}. 

Source intensity spectra at energies above a few tens of GeV from high-redshift populations are attenuated
by interactions with the EBL, which consists of infrared, optical, and ultraviolet photons
primarily from direct and reprocessed starlight throughout cosmic
history. While {\em Fermi} observations have produced
constraints on models of the EBL \cite{fermiebl,markosebl}, the
details remain quite uncertain
\cite{floydebl,chuckebl,Kneiske+Dole_2010,joelebl} .
In some of the models we consider, we use a broken power-law model for
the non-dark-matter emission, which can act as a proxy for EBL
attenuation as well as for accounting for the intrinsic properties of the emitting sources.

We assume an anisotropy for our composite power law consistent with
observations rather than tying it to models of a given source
class. A significant contribution to the anisotropy is expected to
come from unresolved blazars. The anisotropy properties of the blazar
contribution to the background are generally dependent on the details
of the blazar luminosity function, and therefore similarly uncertain
as the overall amplitude of the collective blazar intensity. The
derived anisotropy of the IGRB is comparable to the level of blazar
fluctuation anisotropy predicted by theoretical work \citep[see,
e.g.,][]{Ando:2006cr}, though blazars are expected to be too anisotropic to contribute the entire IGRB~\citep{Cuoco:2012yf, Harding:2012gk}.

For the dark matter component in our example scenarios, we model the emission from pair
annihilation of WIMP dark matter particles in Galactic subhalos.  We
consider the photon intensity spectra given in~\citep{Fornengo:2004kj}
produced by annihilation into two benchmark final states: (1)
$b\bar{b}$, which generates a relatively soft continuum photon
spectrum primarily from the decay of neutral pions produced by the
hadronisation of quark jets, and (2) $\tau^{+}\tau^{-}$, which
produces a harder photon spectrum due to a significant contribution
from final state radiation associated with the production of charged
leptons.  We choose values of the annihilation cross section between
the canonical value for a thermal relic $\langle \sigma v\rangle_{0} =
3.0 \times 10^{26} \rm \, cm^3 s^{-1}$~\cite{Jungman:1995df}
\citep[see also][]{Steigman:2012nb} and 33
times that value. Different constraints on dark matter models can be
obtained under different assumptions for various targets,
\citep[e.g.,][]{Ackermann:2011wa, GeringerSameth:2011iw,
  Cholis:2012am, Mazziotta:2012ux, Ackermann:2012rg, Hooper:2011ti,
  Hooper:2012sr}. The fluctuation anisotropy from dark matter
annihilation, which is determined exclusively by the spatial
distribution of the dark matter, has been predicted for annihilation
in Galactic dark matter subhalos by \citet{SiegalGaskins:2008ge,
  Fornasa:2009qh, Ando:2009fp, Fornasa:2012gu}. We present three
models with a dark matter component: one that falls within the typical predictions for the dark matter anisotropy and two models that do not.  However, we emphasise that the model parameters adopted for each example scenario were chosen to be illustrative of the
decomposition methods rather than to represent the most plausible compositions
of the \emph{Fermi}-LAT IGRB\@. 

\subsection{Error Analysis}
We compute error bars for the example IGRB intensity and anisotropy energy spectra assuming observations with the \emph{Fermi}-LAT\@.
The 1$\sigma$ error bars for the total fluctuation anisotropy in each energy bin were
computed using the formula \citep{Knox:1995dq}:
\beq
\ \Delta \hat{C}_\ell = \sqrt{\frac{2}{\left(2\ell +
      1\right)\Delta\ell f_{\mathrm sky}}}\left(\hat{C}_\ell + \frac{\hat{C}_{\rm N}}{W_\ell^2}\right)
\eeq
where $\hat{C}_{\ell}$ is the total fluctuation angular power
spectrum, $\Delta\ell$ is the width of the multipole bin, $f_{\rm
  sky}$ is the fraction of the sky used to calculate the angular power
spectrum, $\hat{C}_{\rm N} = \left(4\pi f_{\rm sky} /
  N_{\gamma}\right)$ is the power spectrum of the photon noise
associated with the total measured emission, with $N_{\gamma}$ the
total number of photons collected during the observation period in the
sky region analyzed, and $W_\ell$ is the beam window function of the instrument.  

We approximate the PSF of the LAT as a circular Gaussian beam with energy-dependent width $\sigma_\mathrm{b}(E)$, determined from the 68\% containment angle radius reported in the P7\_V6 performance curves~\footnote{{\tt http://www.slac.stanford.edu/exp/glast/groups/canda/\\lat\_Performance.htm}}, so $W_\ell =
\mathrm{exp}\left(-\ell^2\sigma_\mathrm{b}^2/2\right)$, which is the window function of a
Gaussian beam of width $\sigma_\mathrm{b}$. We evaluate $\sigma_\mathrm{b}$ at the log center of the energy bin.  For the example scenarios, we show anisotropy energy spectra at $\ell=175$, and take $\Delta\ell = 50$, choices made to ease comparison with the results reported in the \emph{Fermi} anisotropy analysis~\citep{Ackermann:2012uf} which used $\Delta\ell = 50$ and focused on the Poisson angular power measured at $\ell \gtrsim 150$ to limit contamination from Galactic diffuse emission.

Following the \emph{Fermi} anisotropy analysis, we assume a sky fraction $f_{\rm sky} =
0.32$ is used to perform the anisotropy and intensity measurements, i.e., a large fraction of the sky is masked.  We take the field of view of the LAT to be $\Omega = 2.4$ sr, and approximate the energy-dependent effective area of the LAT from the reported performance
curves.  For a specified all-sky observation time $t_{\rm obs}$, we
calculate the number of photons detected outside the mask to be
$N_{\gamma}= \int_{E_{\rm min}}^{E_{\rm max}} {\rm d}E\,\, \frac{{\rm
    d}I}{{\rm d}E}\, A_{\rm eff}(E)\, \Omega\, f_{\rm sky}\, t_{\rm obs}$,
where $\frac{{\rm d}I}{{\rm d}E}$ is the total (energy-dependent)
differential intensity of the IGRB, $A_{\rm eff}$ is the energy-dependent effective
area of the instrument, and we have assumed observations
in all-sky survey mode and uniform sky exposure.

The errors on the intensity energy spectra represent the Poisson noise associated with the number of photons collected in each energy bin, as well as an assumed 20\% uncertainty on the effective
area of the instrument as estimated from the performance curves; we
note that the uncertainty on the effective area does not contribute to
the uncertainty in the fluctuation angular power spectra since
normalising intensity fluctuations to the mean map intensity removes
the effective area from the calculation.  For the all-sky observation
time of 10 years assumed in the examples presented in this work, we find that for the adopted energy binning, the
uncertainty in $A_{\rm eff}$ dominates the error bars on the intensity spectra up to a few hundred GeV
for all of the scenarios considered here, resulting in relatively uniform
error bars as a function of energy.

The errors on the simulated intensity and anisotropy spectra can be
propagated through the decomposition equations, enabling calculation
of error bars for the decomposed intensity spectra. Often, the $\hat{C}_{\ell}$ of one source class must be obtained
from a flat baseline in the anisotropy energy spectrum over the
energies where that source class dominates the anisotropy. An estimate for the
$\hat{C}_{\ell}$ of such a source class can be
obtained by taking the weighted mean of the baseline points with error
equal to the weighted error of the mean. The decomposed intensity
spectra can then be written in terms of
quantities for which a mean and error bar can be computed. The
1$\sigma$ and 3$\sigma$ confidence intervals are determined via Monte
Carlo.

Systematic errors due, for instance, to a non-Gaussian PSF, or uncertainties in foreground subtraction, are assumed to be subdominant to the
statistical uncertainties and therefore negligible. After ten years of
observations, we expect that our understanding of both the instrument
and the gamma-ray sky to have significantly improved. While this may
lead to optimistic predictions of \emph{Fermi}'s 10-year capabilities,
we stress that our primary goal is to demonstrate a suite of
decompositon techniques rather than make detailed predictions of
upcoming IGRB observations.  However, the
systematic uncertainties will have to be carefully considered when
applying our techniques to actual data.

\section{Two-Component Decomposition: Applications}
\label{sec:details}

The formalism for each of the six two-component decomposition techniques summarised in Table~\ref{table1} is given in the following subsections.  

\subsection{Double Plateau} 
\begin{figure*}
 \centering
\includegraphics[width=0.45\textwidth]{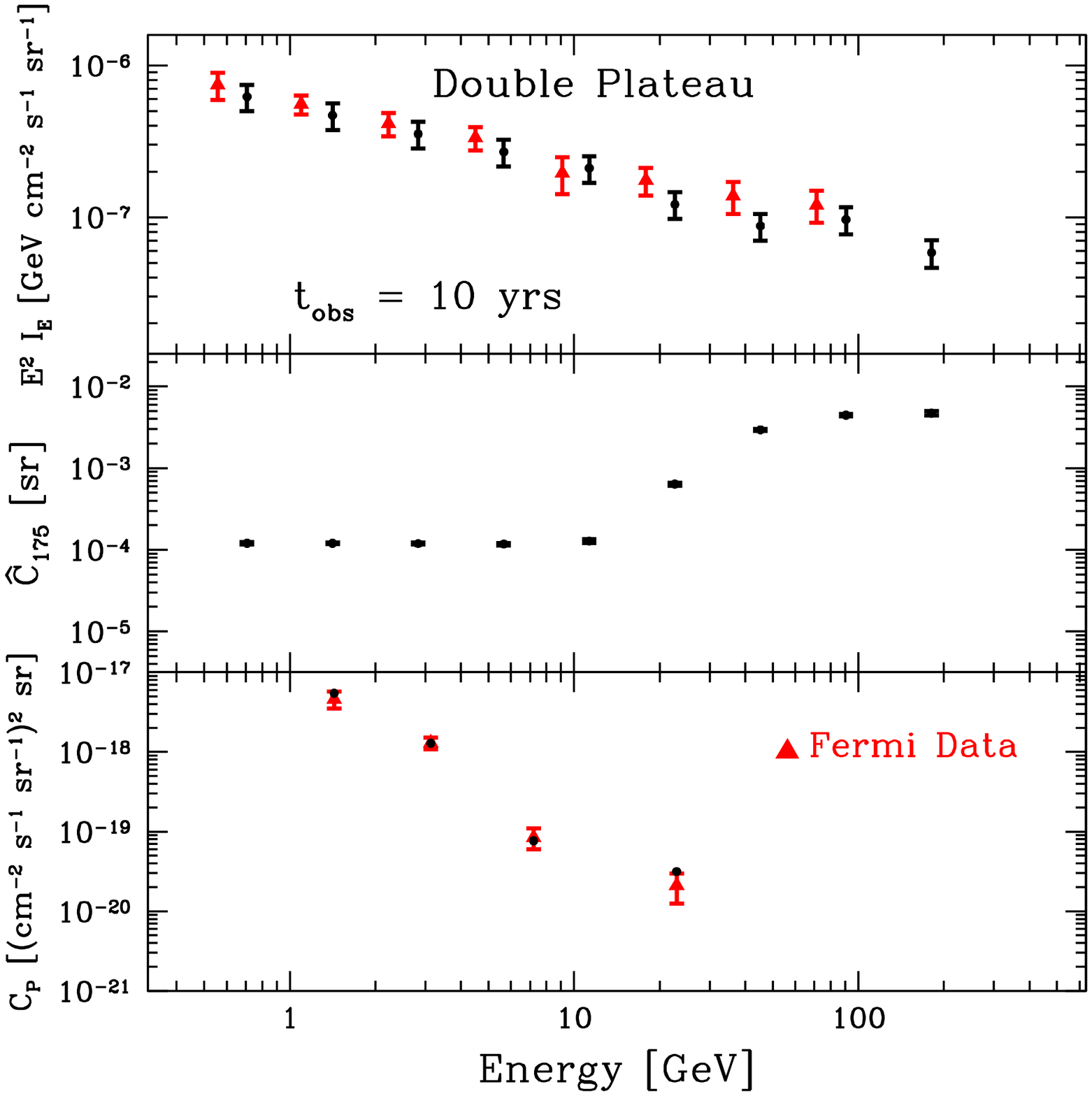}
\includegraphics[width=0.45\textwidth]{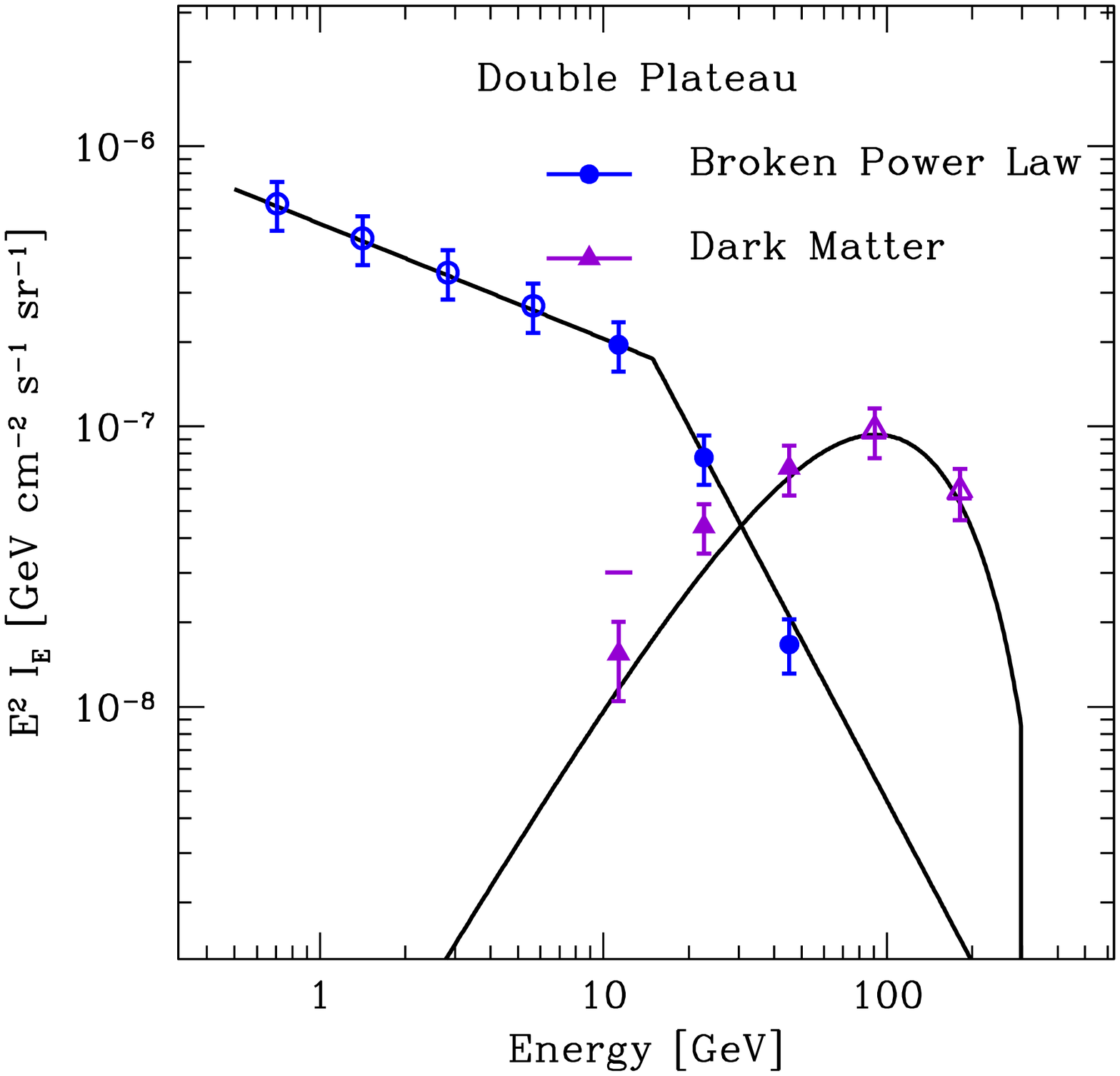}
\caption{Example double plateau decomposition.  \emph{Left:} Total IGRB
  intensity (top),  anisotropy energy spectrum (middle), and angular
  power $C_P$ (bottom). Energy bins are equally spaced in log space,
  and all quantities are reported at the log center of the energy
  bin. Error bars
  assume $t_{\rm obs}=10$ years of \emph{Fermi}-LAT observations in sky-survey mode. If a data point is within $3\sigma$ of zero, we
  place a $3\sigma$ upper limit bar in addition to the $1\sigma$
  error bars. Red triangles indicate the \emph{Fermi} IGRB intensity
  measurements from~\citep{Abdo:2010nz:igrb} for ten months of observation,
  and angular power measurements from~\citep{Ackermann:2012uf} for 22
  months of observation. The dark matter intensity
  spectrum corresponds to a $m_{\rm DM}=300$~GeV particle annihilating
  to $\tau^{+}\tau^{-}$. \emph{Right:} The decomposed intensity energy
  spectrum of the power law component (blue circles) and dark matter annihilation
  (purple triangles) recovered using the plateau technique. The baseline points (open plot symbols) from which $\cla$ and $\clb$ were determined were not decomposed. Each component's input intensity spectrum is overlaid in black. \label{fig:plateau}}  
\end{figure*}

As previously described, we can infer the value of $\cla$ by observing a flat baseline (plateau) in the anisotropy energy spectrum at either low or high energies. In the event that
we observe a second such plateau in the anisotropy energy spectrum
(Fig.~\ref{fig:plateau}) with amplitude either above or below the level of the first plateau
$\cla$ ({\it i.e.} where $\ina \ll \inb$), 
we can obtain a value for $\clb$.
In this case
Eqs.~\ref{Ieg} and \ref{Idm} for the intensity spectra of the two components can be solved directly.  
This is a \emph{double plateau decomposition}, and corresponds to the
case that one of the components dominates the anisotropy at low
energies, while the other dominates the anisotropy at high energies.
In this case we assume the anisotropy energy spectrum increases or
decreases monotonically between the two plateaus; the case of local extrema is discussed in the case of a minimum decomposition in~\S\ref{sec:min}.  A double plateau is a
particularly ideal case because the input spectra can be derived
exactly, without making any assumptions about the relative
contributions of the source classes to either the total intensity or
anisotropy that cannot be inferred directly from the observed spectra.

As an example scenario, shown in Fig.~\ref{fig:plateau}, we choose $m_{\rm DM}$ = 300 GeV,
$\langle \sigma v \rangle$ =
20$\langle \sigma v \rangle_{0}$, and annihilation into $\tau^{+}\tau^{-}$ for
the dark matter intensity spectrum, and adopt a broken power law for
the remaining intensity spectrum. The
anisotropies were taken to be $\hat{C}_\ell = 1.2 \times 10^{-4}$~sr,
and $\hat{C}_\ell = 5 \times 10^{-3}$~sr for the broken power law and dark
matter signals, respectively. Because the anisotropy energy spectrum
is still rising slightly between the last two data points, the estimate for the
$\hat{C}_\ell$ of the dark matter component will be biased low, thus slightly
biasing the decomposed spectra away from the true value.

%%%%%%%%%%%%%%%%%%%%%%%%%%%%%%%%%%%%%%%%%%%%%%%%%%%%%%%%%%%%%%%%%%%%%%%%%%%
% Plateau Figure
%%%%%%%%%%%%%%%%%%%%%%%%%%%%%%%%%%%%%%%%%%%%%%%%%%%%%%%%%%%%%%%%%%%%%%%%%%%

\subsection{Low-Anisotropy Plateau} 
\label{sec:risinganiso}
We now consider a scenario in which a low-anisotropy plateau is measured in the anisotropy energy spectrum at low or high energies, and the anisotropy rises from the
plateau at low energies or falls to the plateau at high energies.
This corresponds to a case where the component subdominant in
intensity at the plateau has a much higher anisotropy, 
\begin{equation}
 \cla \ll \clb = \Lambda \cla
\end{equation}
with $\Lambda \gg 1$. 

In this case, Eq.~(\ref{two}) can be written as
\begin{eqnarray}
\clo 
& = &  \left(1-\frac{\inb}{\ino}\right)^2\cla + \left(\frac{\inb}{\ino}\right)^2\Lambda \cla 
\end{eqnarray}
or
%, by dividing out by $\cla$ and writing out the first square, 
\begin{equation} 
\frac{\clo}{\cla} = 1 -2\frac{\inb}{\ino} + (1+\Lambda)\left(\frac{\inb}{\ino}\right)^2 .
\end{equation}
Now we define $x(E)=\inb/\ino$ and $\omega (E)= \clo/\cla -1$, which can be determined by observations at each energy. Then we have
\begin{equation}
(1+\Lambda)x^2 -2x - \omega = 0
\end{equation}
with solution 
\begin{equation}
x = \frac{1\pm\sqrt{1+(1+\Lambda)\omega}}{1+\Lambda} .
\end{equation}
Since $\Lambda \gg 1$, as long as $\omega > 1$ we can approximate this by 
\begin{equation}
x \approx \frac{1\pm\sqrt{(1+\Lambda)\omega}}{1+\Lambda} \approx \frac{\sqrt{\omega}}{\sqrt{1+\Lambda}}\,.
\end{equation} 
where we have selected the $+$ solution since $x$ is a non-negative quantity. Since $\omega$ is an
observable, it is always possible to determine whether the $\omega >
1$ condition holds. The shape of the subdominant
spectrum can thus be derived up to a multiplicative constant.

\begin{figure*}
 \centering
  		\scalebox{0.40}{\includegraphics{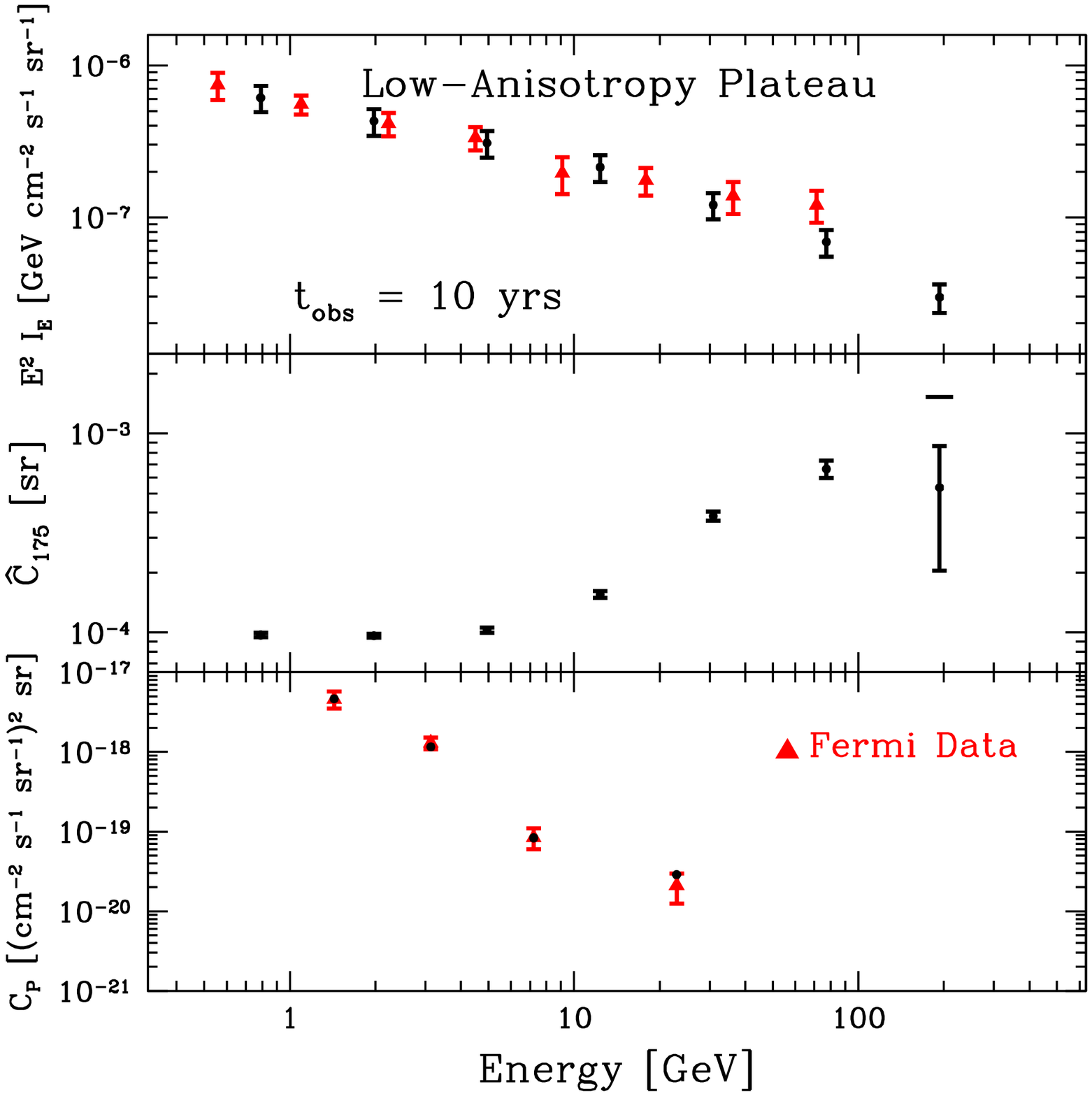}}
  		\scalebox{0.40}{\includegraphics{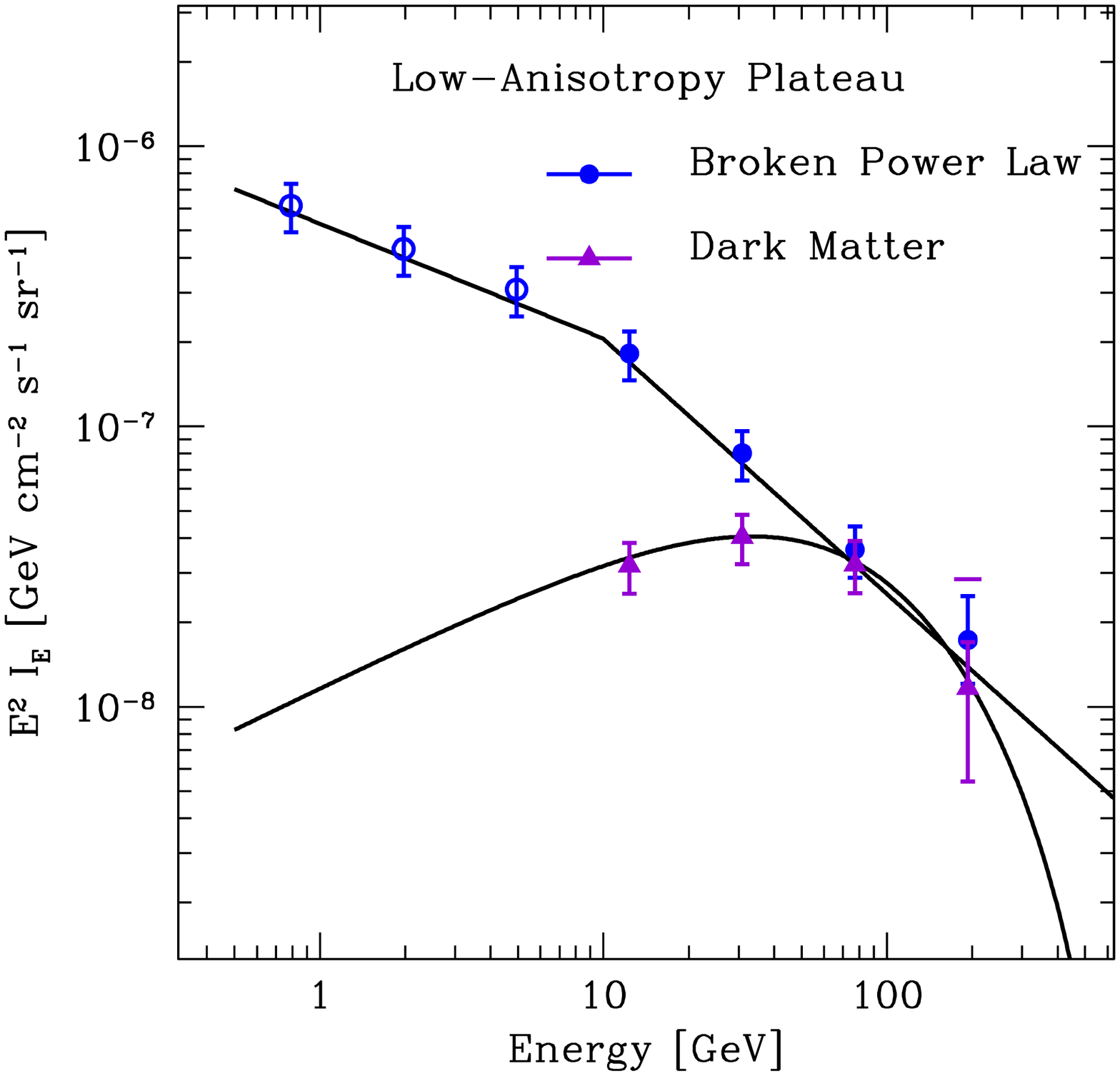}}
\caption{Example low-anisotropy plateau decomposition.  \emph{Left:} Total IGRB
  intensity (top),  anisotropy energy spectrum (middle), and angular
  power $C_P$ (bottom). Energy bins are equally spaced in log space,
  and all quantities are reported at the log center of the energy
  bin. Error bars
  assume $t_{\rm obs}=10$ years of \emph{Fermi}-LAT observations in
  sky-survey mode. If a data point is within $3\sigma$ of zero, we
  place a $3\sigma$ upper limit bar in addition to the $1\sigma$
  error bars. In this scenario the IGRB is composed of emission from
  a broken power law component and Galactic dark matter annihilation. Red triangles indicate the \emph{Fermi} IGRB intensity
  measurements from~\citep{Abdo:2010nz:igrb} for ten months of observation,
  and angular power measurements from~\citep{Ackermann:2012uf} for 22
  months of observation.  The dark matter intensity
  spectrum corresponds to a $m_{\rm DM} = 1000$~GeV particle annihilating
  to $b\bar{b}$. \emph{Right:} The decomposed intensity energy
  spectra of the power law component  (blue circles) and dark matter annihilation
  (purple triangles) are recovered using the low-anisotropy plateau technique. The baseline points at low energy (open plot symbols) from which $\cla$ was determined were not decomposed. Each component's input intensity spectrum is overlaid in black. Note that this method recovers the normalisations of each component's intensity spectrum up to a multiplicative constant; the constant has been set to the true value to facilitate comparison with the input spectral shapes.\label{fig:lo_plat}}
\end{figure*}

As an example scenario, shown in Fig.~\ref{fig:lo_plat}, we choose a dark matter particle with $m_{\rm DM} = 1000$~GeV that annihilates to $b\bar{b}$ with $\langle \sigma v \rangle$ =
33$\langle \sigma v \rangle_{0}$ and a broken power law component to the
intensity spectrum.  We set the fluctuation angular
power to $\hat{C}_\ell = 1 \times 10^{-4}$~sr for the
broken power law component,
and $\hat{C}_\ell = 2.5 \times 10^{-3}$~sr for dark matter.

%%%%%%%%%%%%%%%%%%%%%%%%%%%%%%%%%%%%%%%%%%%%%%%%%%%%%%%%%%%%%%%%%%%%%%%%%%%
% Low Anisotropy Plateau Figure
%%%%%%%%%%%%%%%%%%%%%%%%%%%%%%%%%%%%%%%%%%%%%%%%%%%%%%%%%%%%%%%%%%%%%%%%%%%

\subsection{High-Anisotropy Plateau} 

If a high-anisotropy plateau is measured in the anisotropy energy spectrum at low or high energies, and the anisotropy falls from the plateau at low energies
or rises to the plateau at high energies, then a
less anisotropic source must be making an increasing contribution to the
background at energies far from the plateau.  We now consider this scenario, corresponding to the case that one component is {\em everywhere} dominant in the intensity, 
\begin{equation}
\ina > \inb
\end{equation}
and also more anisotropic 
\begin{equation}
\cla \gg \clb.
\end{equation}

In this case, Eq.~\ref{two} can be approximated by 
\begin{equation}
\clo \approx  \left(\frac{\ina}{\ino}\right)^2\cla .
\end{equation}
Immediately then we have
\begin{equation}
\ina \approx \ino \sqrt{\frac{\clo}{\cla}}
\label{eq:fall1}
\end{equation}
and 
\begin{equation}
\inb \approx \ino \left(1-\sqrt{\frac{\clo}{\cla}}\right).
\label{eq:fall2}
\end{equation}

 The assumption that the higher anisotropy source is
dominant must be satisfied at the energies where the high-anisotropy plateau is measured. Each subsequent point
moving away from the plateau in energy will yield a value for the fractional contribution
of the subdominant source via the ratio of Eqns.~\ref{eq:fall1} and \ref{eq:fall2}. Therefore, the
appropriateness of the assumption that the anisotropic source is
dominant can always be verified. This decomposition yields both the
shape of the intensity spectra as well as their normalisations.

\begin{figure*}
 \centering
  		\scalebox{0.40}{\includegraphics{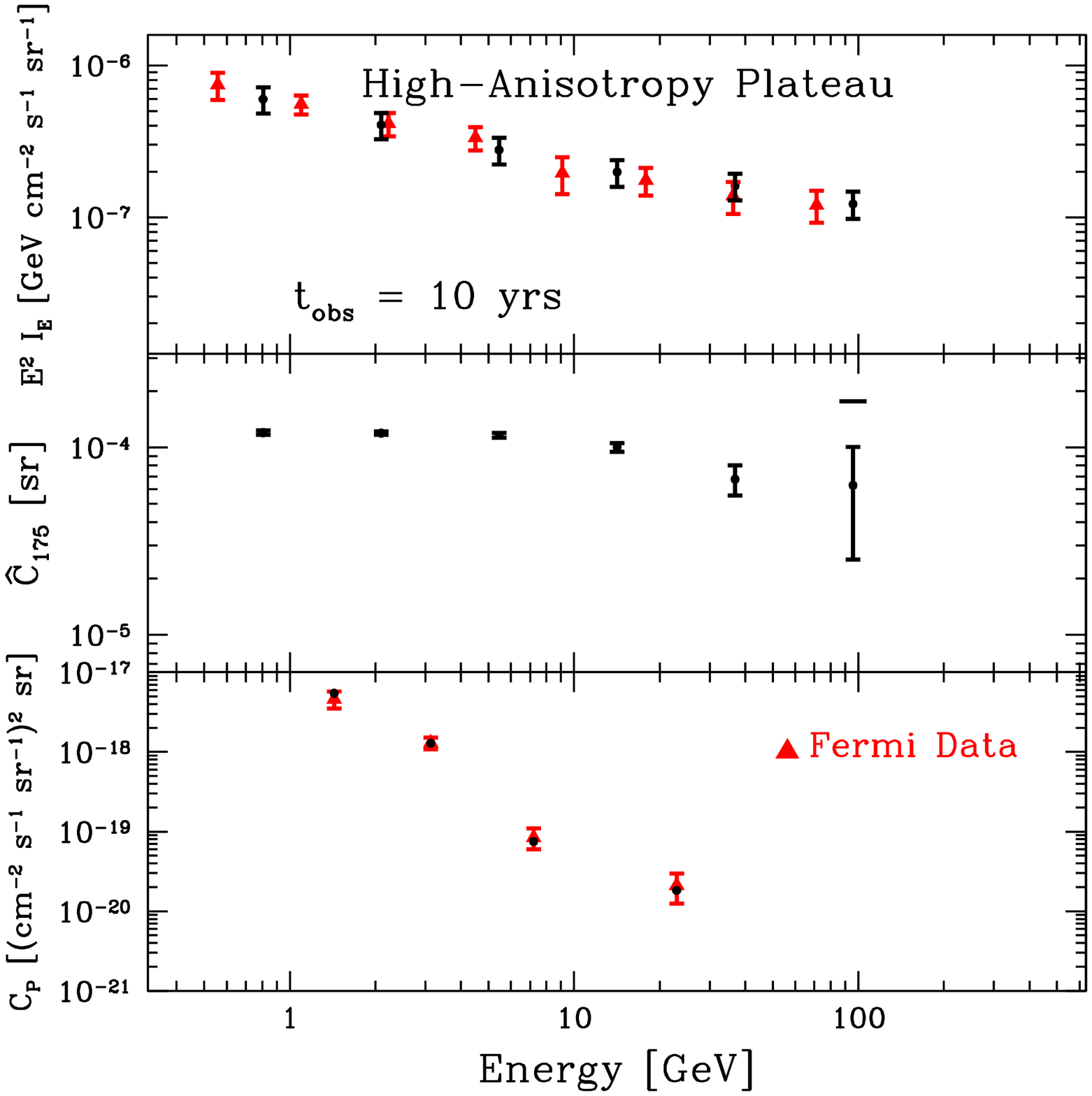}}
  		\scalebox{0.40}{\includegraphics{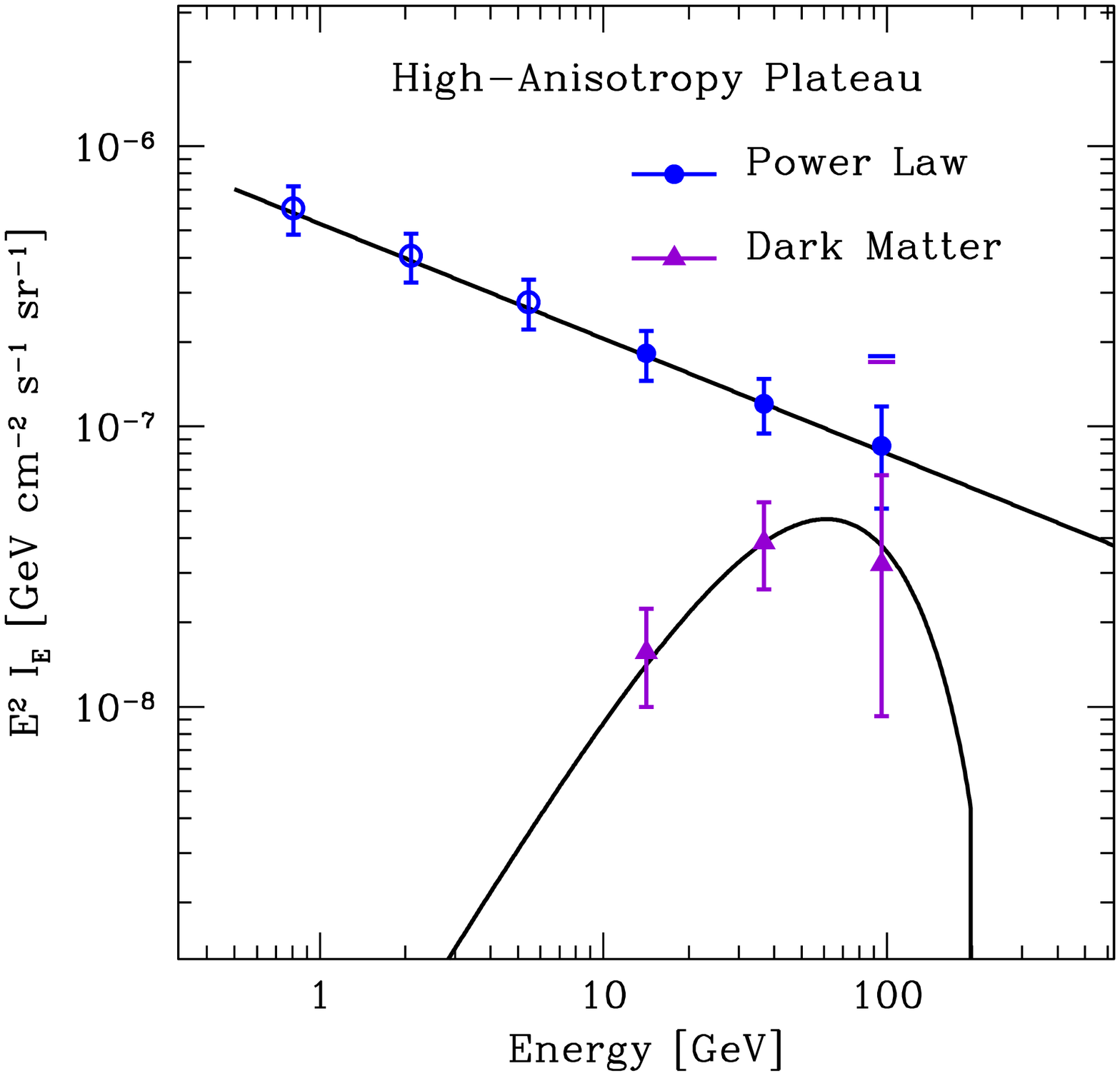}}
\caption{Example high-anisotropy plateau decomposition.  \emph{Left:} Total IGRB
  intensity (top),  anisotropy energy spectrum (middle), and angular
  power $C_P$ (bottom). Energy bins are equally spaced in log space,
  and all quantities are reported at the log center of the energy
  bin. Error bars
  assume $t_{\rm obs}=10$ years of \emph{Fermi}-LAT observations in
  sky-survey mode. If a data point is within $3\sigma$ of zero, we
  place a $3\sigma$ upper limit bar in addition to the $1\sigma$
  error bars. In this scenario the IGRB is composed of emission from
  a power-law component and Galactic dark matter annihilation. The dark matter intensity
  spectrum corresponds to a $m_{\rm DM} = 200$~GeV particle annihilating
  to $\tau^{+}\tau^{-}$. \emph{Right:} The decomposed intensity energy
  spectra of the power law component (blue circles) and dark matter annihilation
  (purple triangles) recovered using the low-anisotropy plateau technique. The baseline points at low energy (open plot symbols) from which $\cla$ was determined were not decomposed. Each component's input intensity spectrum is overlaid in black. Red triangles indicate the \emph{Fermi} IGRB intensity
  measurements from~\citep{Abdo:2010nz:igrb} for ten months of observation,
  and angular power measurements from~\citep{Ackermann:2012uf} for 22
  months of observation.  \label{fig:hi_plat}}
\end{figure*}

As an example scenario,  shown in Fig.~\ref{fig:hi_plat}, we choose a dark matter particle with $m_{\rm DM} = 200$~GeV that annihilates to $\tau^{+}\tau^{-}$ with $\langle \sigma v \rangle$ =
6.7$\langle \sigma v \rangle_{0}$ and a component with a power law
intensity spectrum.  We set the fluctuation angular
power to $\hat{C}_\ell = 1.2 \times 10^{-4}$~sr for
the power law component,
and $\hat{C}_\ell = 5 \times 10^{-6}$~sr for dark
matter. These parameters are consistent with current observational
and theoretical constraints on the IGRB intensity, anisotropy, and
dark matter properties.

%%%%%%%%%%%%%%%%%%%%%%%%%%%%%%%%%%%%%%%%%%%%%%%%%%%%%%%%%%%%%%%%%%%%%%%%%%%
% High Anisotropy Plateau Figure
%%%%%%%%%%%%%%%%%%%%%%%%%%%%%%%%%%%%%%%%%%%%%%%%%%%%%%%%%%%%%%%%%%%%%%%%%%%

\subsection{Known Zero-Anisotropy Component}

If one of the components of a two-component background is completely
isotropic, then  Eq.~\ref{two} becomes:

\beq
\clo = \left(\frac{\ina}{\ino}\right)^2 \cla
\eeq
which may be rewitten simply as
\beq
\ \ina = \ino \sqrt{\clo / \cla}.
\eeq
Thus, in this case the shape of the spectrum of the component with
nonzero anisotropy can be determined up to a
multiplicative constant by measuring $\ino(E)$ and $\clo(E)$.  This
technique is a special case of the high-anisotropy plateau where
Eqns.~\ref{eq:fall1} and~\ref{eq:fall2} hold exactly, regardless of which source is dominant. We emphasise that this method requires {\em a priori} knowledge that a zero-anisotropy component exists, but does not require knowledge of the shape or normalisation of its intensity spectrum.

\begin{figure}
 \centering
  		\scalebox{0.40}{\includegraphics{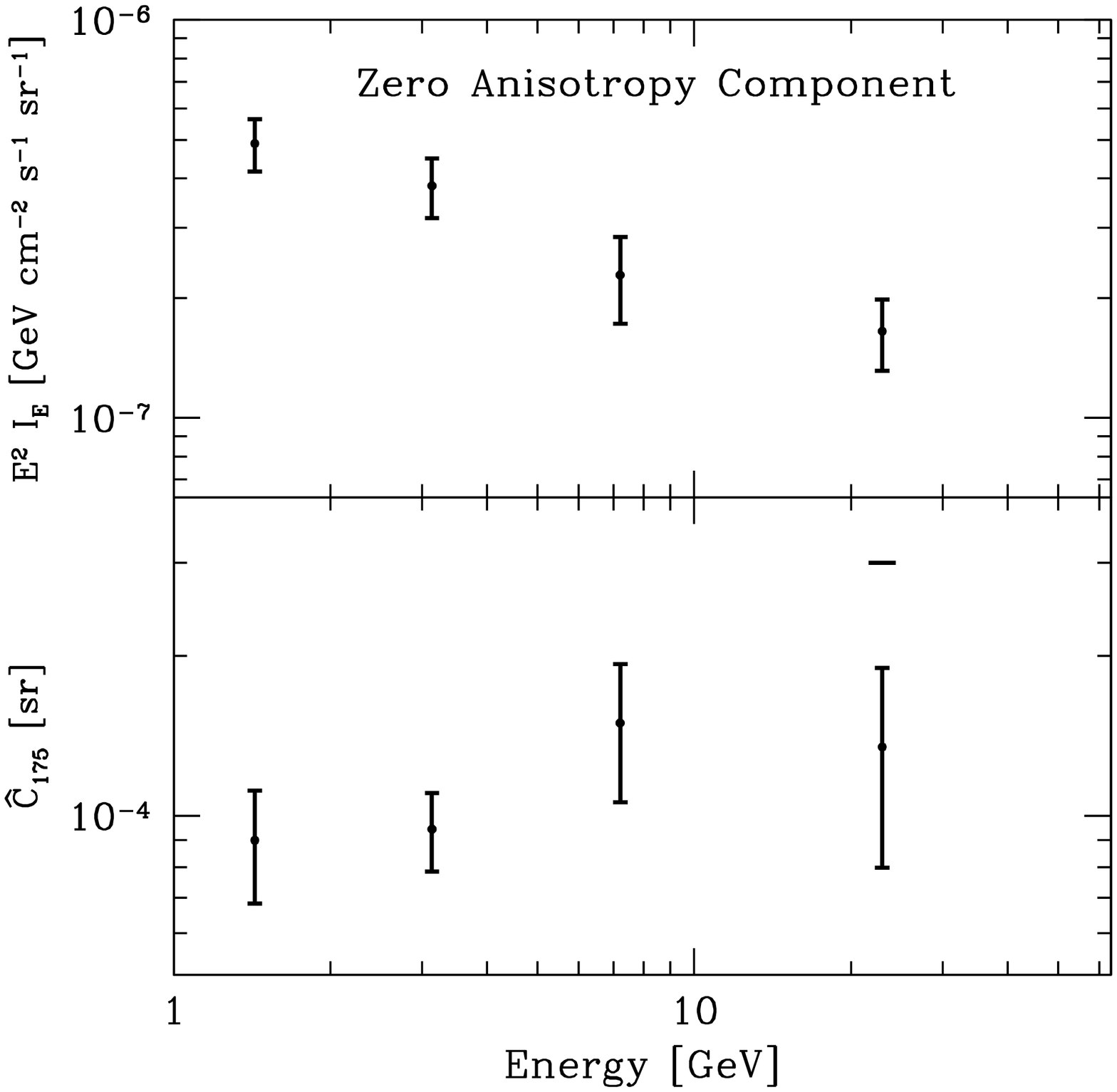}}
		\scalebox{0.40}{\includegraphics{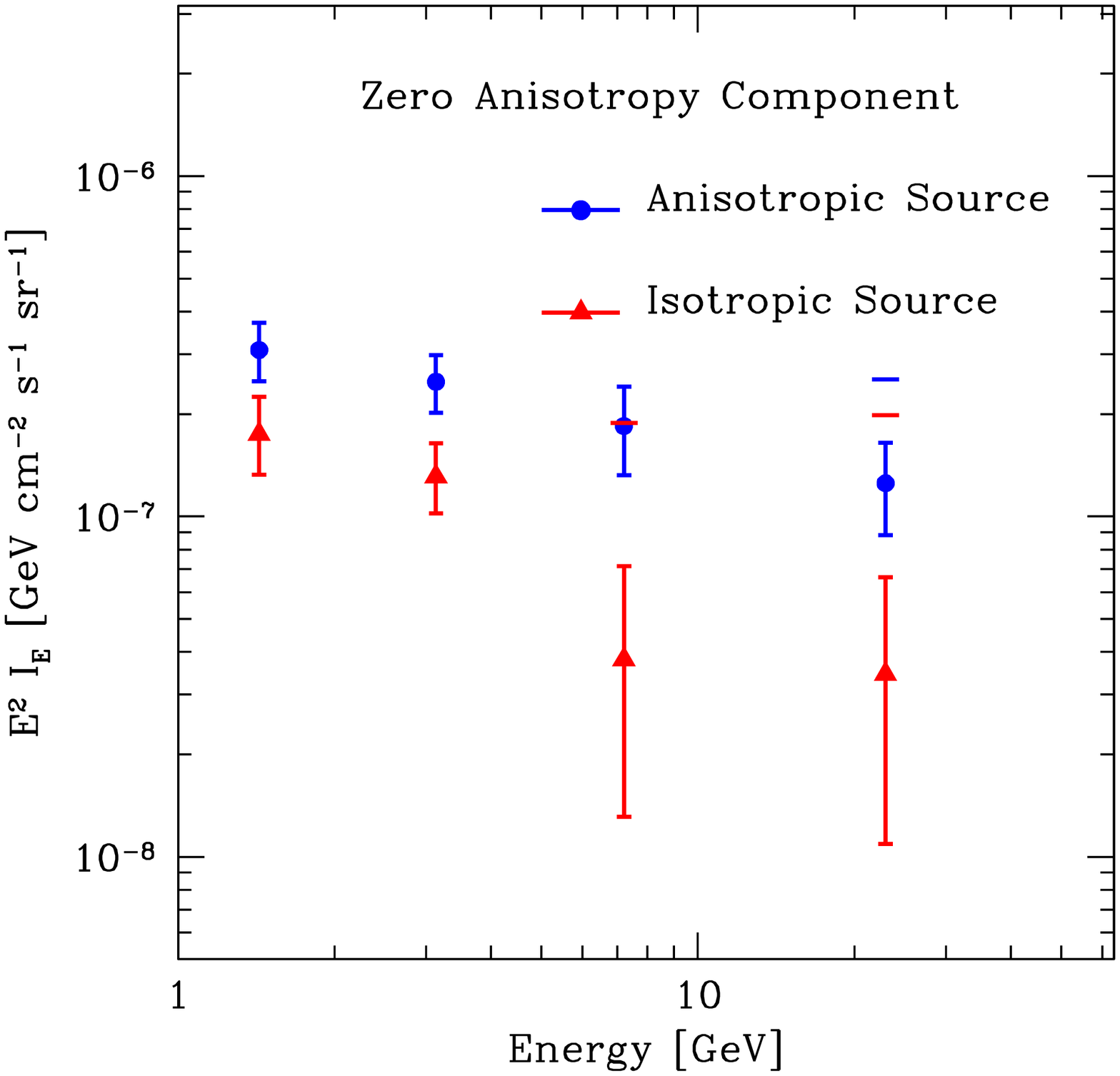}}
\caption{Example known zero-anisotropy component decomposition.  \emph{Top:} Total IGRB
  intensity (top) and anisotropy (bottom) energy spectra as measured
  by \emph{Fermi}-LAT using 11 months of data for the intensity points
  and $22$~months of data for the
  anisotropy points. If the data point is within $3\sigma$ of zero, we
  place a $3\sigma$ upper limit bar in addition to the $1\sigma$
  error bars. 
  \emph{Bottom:} The decomposed intensity energy
  spectra of the anisotropic component (blue circles) and the zero-anisotropy component
  (red triangles) are recovered using the known zero-anisotropy
  component technique. Since the normalisation of the intensity
  spectrum of the anisotropic component is not recovered by this
  technique, we choose it to be consistent with a blazar-like source
  class with $\hat{C}_{175} = 2.2 \times 10^{-4}$ sr.
\label{fig:clzero} }
\end{figure}

To demonstrate such a decomposition, as shown in
Fig.~\ref{fig:clzero}, we analyze the intensity and
anisotropy data as measured by \emph{Fermi}-LAT\@; note that the measured anisotropy energy spectrum is consistent with no energy dependence.  

The decomposition presented in
Fig.~\ref{fig:clzero} yields spectra with slope consistent
with the slope of the composite IGRB\@.  This decomposition technique
recovers the shape but not the normalisation of the intensity spectrum
of the anisotropic component, hence the normalisation of its
decomposed intensity spectrum shown in Fig.~\ref{fig:clzero} is arbitrary.  We caution the reader that this
is not a definitive decomposition of the \emph{Fermi}-LAT IGRB due to
uncertainty in matching the intensity and anisotropy measurements from
two different studies as well as the possible presence of other
contributing source classes\@.

%%%%%%%%%%%%%%%%%%%%%%%%%%%%%%%%%%%%%%%%%%%%%%%%%%%%%%%%%%%%%%%%%%%%%%%%%%%
% Zero Anisotropy Figure
%%%%%%%%%%%%%%%%%%%%%%%%%%%%%%%%%%%%%%%%%%%%%%%%%%%%%%%%%%%%%%%%%%%%%%%%%%%

\subsection{Minimum} 
\label{sec:min}
Assuming $\cla$ can be inferred from the data by measuring a plateau
in the anisotropy energy spectrum, an additional way to determine
$\clb$ and decompose the observed intensity spectrum exists if a local
minimum 
is observed in the anisotropy energy spectrum. The condition for observing a minimum in the anisotropy energy 
spectrum can be obtained by differentiating Eq.~(\ref{two}) with respect to energy. We find that a minimum occurs when
\beqar
\clb \left(\frac{\ina(E_{\rm min})}{\ino(E_{\rm min})} - 1 \right) +
\cla \frac{\ina(E_{\rm min})}{\ino(E_{\rm min})} = 0\,. \label{eq:Min}
\eeqar
At the energy at which the local minimum occurs $E_{\rm min}$, we can simultaneously solve this equation with Eq.~\ref{one} at the same energy to obtain
\beqar
\ \clb = \frac{\cla \clo(E_{\rm min})}{\cla - \clo(E_{\rm min})} \label{eq:Cldm}.
\eeqar
Substituting this relation into Eqs.~\ref{Ieg} and \ref{Idm}
completely determines the intensity energy spectrum for both source classes.
In this case, a {\em minimum decomposition} is possible. For
appropriate levels of anisotropy and fractional contribution to the
background, Eq.~\ref{two} states that the total anisotropy can be less than the anisotropy of either
source class.  Indeed,
the minimum will exist only if there is an energy at which $\ina \cla
= \inb \clb$, and this requirement will always produce a minimum rather
than a maximum in the total anisotropy since

\beq
\ \frac{d^2 \clo}{d E^2} \bigg|_{E_{\rm min}} = 2 \left(\frac{d}{dE}
  \frac{\ina}{\ino}\right)^2 \left(\cla + \clb\right)
\eeq
is non-negative. It also implies that a component which is
always subdominant cannot be decomposed using this technique if its
angular power is also smaller than that of the dominant component. A
second caveat of this method is that a minimum can also occur when
$\frac{d}{dE}\left(\frac{\inb}{\ino}\right) = 0$, which does not yield
a constraint on $\clb$. This
occurs when a low anisotropy component transitions from contributing
an increasing fraction of the total intensity to contributing a
decreasing fraction of the total intensity. Thus, the anisotropy of
the resulting minimum must be intermediate between the anisotropy of
each source class. On the other hand, the decomposition minimum occurs at an anisotropy below that of either source class. Hence, in practice, the two types of minima may be distinguished if the anisotropy energy spectrum is observed to take on values both above and below an observed baseline, in which case it must be a decomposition minimum.

The biggest challenge in the applicability of this technique is that
the uncertainties associated with an observed anisotropy energy
spectrum will often be too high to allow a minimum to be measured. The
depth of the minimum is given by the ratio of $\clo$ at the minimum
and the lesser of $\cla$ and $\clb$. By
rearranging Eq.~\ref{eq:Cldm} and assuming $\cla > \clb$ we obtain 

\beqar
\ \frac{\clo}{\clb} = \frac{\cla}{\cla + \clb}
\eeqar 
Thus the depth of the minimum varies between $0.5$ and $1$ times $\cla$, with the largest depth occurring for $\cla=\clb$. For instance, although the example given in \S\ref{sec:risinganiso} (low-anisotropy plateau) in principle exhibits a local minimum in the anisotropy energy spectrum, the depth of this minimum would be very small and thus challenging to measure observationally.

However, this does not mean that likely two-component scenarios for
the IGRB that would allow a minimum decomposition do not exist. One
such scenario can be envisioned if blazars typically exhibit spectral breaks. In combination with the different population-average spectral indices of BL Lacs and FSRQs, this could lead to a scenario in which over a certain energy range the IGRB is composed of emission from two source classes (BL Lacs and FSRQs), and the dominant contributor to the intensity transitions between the source classes at a few GeV (see, e.g., Fig.~3 of \citet{tonivaso11}).  If BL Lacs and FSRQs have approximately equal levels of fluctuation angular power (not unlikely, as \emph{Fermi} has resolved comparable numbers of sources in each of the two classes), the minimum in the anisotropy energy spectrum would appear around the transition energy of a few GeV, where photon statistics would be  fairly large and thus may allow a measurement of the minimum.

We note that by examining the conditions under which a minimum occurs, we can also better understand
the degeneracy in choice of sign in Eqs.~\ref{Ieg} and \ref{Idm}.  By rearranging Eq.~\ref{Idm}, which gives
\beqar
\inb \clb - \ina \cla = \\ \pm \ino \sqrt{\cla \clo + \clb \clo - \cla \clb} \nonumber
\eeqar
we see that the proper sign is determined based upon the sign of $\inb \clb - \ina \cla$. As the relative contributions of the two source classes fluctuate, however, this quantity can go from positive to negative, forcing us to switch the choice of signs when this happens. Notably, a sign change would have to occur when $\ina \cla = \inb \clb$, precisely when $\clo$ is minimised. Indeed, since $\inb/\ino$ is changing through the minimum, the sign \emph{must} change at every minimum with the only exception being when $\inb/\ino$ is simultaneously minimised or maximised, which does not happen in general. Consequently, no single choice of signs describes the entire spectrum, but only the region between two consecutive minima.

\subsection{Decompositions from Multiple $\ell$ s}
In the case where two distinct anisotropy energy spectra, each
measured at a different $\ell$, can be obtained, a full decomposition
is possible provided $\hat{C}_\ell$ varies with $\ell$ for at least one source
class and $\cla$ can be determined for each spectrum, e.g. via a plateau. We still assume $\hat{C}_\ell$  is independent of energy. Differentiating Eq.~\ref{two} with respect to energy and rearranging, we obtain,

\beqar
\ \frac{d}{dE}\left(\frac{\ina}{\ino}\right) = \frac{d \clo(E, \ell) / dE}{2 \{\frac{\ina}{\ino} [\cla(\ell) + \clb(\ell)] - \clb(\ell)\}}
\eeqar
Since the left-hand side is independent of $\ell$, we must have for two different $\ell$ values $\ell_1$ and $\ell_2$ at \emph{any} energy E

\beqar
\ \frac{d \cloo / dE}{2 [\frac{\ina}{\ino} (\clao + \clbo) - \clbo]} = \\
\frac{d \clot / dE}{2 [\frac{\ina}{\ino} (\clat + \clbt) - \clbt]}
. \nonumber 
\eeqar

Using Eq.~\ref{two} to eliminate $\clbo$ and $\clbt$
and solving for $\ina$ yields

\beqar
\ \ina = \ino \frac{\cloo \frac{d \clot}{dE} - \clot \frac{d \cloo}{dE}}{\clao \frac{d \clot}{dE} - \clat \frac{d \cloo}{dE}} .
\eeqar  

This method is particularly ideal in that the input spectra are
derived exactly and without making any assumptions about their relative
intensities or anisotropies. However, this method may be difficult to
implement for the IGRB because we expect most gamma-ray source
populations to produce fluctuation angular power spectra that are
dominated by the Poisson angular power, which takes the same value at
all $\ell$~\citep[e.g.,][]{Ando:2006cr,
  miniati_koushiappas_di-matteo_07, Ando:2009nk,
  SiegalGaskins:2010mp}, with the notable exception of dark matter
annihilation or decay \citep[e.g.,][]{ando_komatsu_06,Ando:2006cr, Cuoco:2010jb,
Fornasa:2012gu}.

\section{Three Component Decomposition}
\label{sec:threecomp}
The separability of the equations for the total intensity and
anisotropy at a given energy into contributions from each component
source class enables a background of multiple components to be decomposed
provided all source classes are uncorrelated and the contributions of
all but two source classes are known.  Furthermore, under certain conditions a decomposition can be performed even when information about additional source classes is more limited.

In this section, we discuss specifically the case of a three-component background. 
The intensity of a three component background as a function of energy is given by
\begin{equation}
\ino = \ina + \inb + \inc
\end{equation}
and the fluctuation anisotropy as a function of energy by
\begin{equation}
\clo(E) = \left(\frac{\ina}{\ino}\right)^2\cla +
\left(\frac{\inb}{\ino}\right)^2\clb + \left(\frac{\inc}{\ino}\right)^2\clc
\end{equation}
again assuming uncorrelated components. 

In the following, we examine the applicability of the techniques we
discussed in two plausible scenarios for the IGRB composition.

\subsection{One component with known intensity and anisotropy}

It is possible that we can obtain, via some other analysis, expressions for both the intensity and anisotropy of a third component, $\inc$ and
$\clc$ as functions of energy. Such a situation could occur for a component arising 
from a population of bright point sources, for which enough individual members
have been resolved so as to obtain a thorough understanding of the spectral behavior, and 
to  constrain well the distribution of source fluxes, $dN/dF$, down to a point from which a reasonable 
extrapolation to even lower fluxes is possible. One can envision, for example, this to be the 
situation for gamma-ray blazars after the completion of the \emph{Fermi} mission. From $dN/dF$ both the anisotropy level 
and the overall intensity normalisation can be calculated, and from the understanding of individual 
source spectra the energy dependence of the intensity can be evaluated. 

We can thus rewrite our equations as
\begin{equation}
\ino - \inc = \ina + \inb
\end{equation}
and
\begin{equation}\label{CLO}
\clo - \left(\frac{\inc}{\ino}\right)^2\clc = \left(\frac{\ina}{\ino}\right)^2\cla + \left(\frac{\inb}{\ino}\right)^2\clb\,,
\end{equation}
where the left hand side of both equations are determinable directly from observables and knowledge of the third component's properties. Hence, we have reduced the problem back to the two-component case where all of our decomposition methods apply.

\subsection{One component with zero anisotropy, and a second component with known spectral shape}

A second likely three-component scenario that is workable in this formalism is one in which no component is completely known, 
but where one component (component~1) has a well-understood and zero (or negligible) anisotropy, and a second component (component~2) 
has a well-understood 
intensity spectral {\em shape}, even if its overall intensity normalisation is unknown.

In the context of the IGRB, the zero-anisotropy component 
could be contamination from unrejected cosmic-ray electrons entering the detector \citep{Abdo:2010nz:igrb} 
or a combination of such cosmic-ray contamination and a very-low anisotropy cosmic component 
(e.g., star-forming galaxies, or cascade emission in the case of significant intergalactic magnetic field \citep{Venters:2012bx}). Blazars, on the other hand, could be the component with a well-understood spectral shape
(since, by the end of the \emph{Fermi} mission, thousands of blazars will have been resolved and have their individual spectra measured), 
even if the overall normalisation of their intensity contribution to the IGRB is still uncertain. 

We will see that in this case the energy dependence of the intensity of a third component can be determined up to a normalisation 
constant. Such information could have extremely high impact if the third component (component 3) 
is, for example, a contribution from dark matter annihilation or decay, as we discuss in \S\ref{sec:disc}. 

Because component~1 has zero anisotropy, Eq.~(\ref{CLO}) becomes 
\begin{equation}\label{CLOzero}
\clo(E) =  \left(\frac{\inb(E)}{\ino(E)}\right)^2\clb + \left(\frac{\inc(E)}{\ino(E)}\right)^2\clc
\end{equation}
where we have written explicitly all energy dependencies. 

We assume, as before, that there is an energy $E_0$ where we know that component 3 does not contribute significantly (as could be the case at low energies for certain 
dark matter annihilation or decay components). At $E_0$, Eq.~(\ref{CLOzero}) then becomes
\begin{equation}\label{CLOE0}
\ino^2(E_0) \clo(E_0) =  \inb^2(E_0)\clb\,.
\end{equation}
Now since component~2 is assumed to have a known spectral shape, we can write
\begin{equation}\label{otherone}
I_2(E) = I_2(E_0)g(E)
\end{equation}
where $g(E)$ is a known function of energy, normalised so that $g(E_0) = 1$. Solving Eq.~(\ref{CLOzero}) for $I_3(E)$ 
and using Eqs. (\ref{CLOE0}) and (\ref{otherone}) to eliminate $\clb$ and $I_2(E)$,
we then obtain
\begin{eqnarray}\label{CLOsolution}
\inc^2(E)\clc
& = & \ino^2(E)\clo(E) - g^2(E) \clo(E_0) \ino^2(E_0)\,. \nonumber \\
\end{eqnarray}

Since all quantities on the right hand side of Eq.~(\ref{CLOsolution}) are known, it follows that $\inc(E)$ can be derived up to a normalisation constant
$1/\sqrt{\clc}$: 
\begin{equation}
\inc(E) =  \sqrt{\frac{\ino^2(E)\clo(E) - g^2(E) \clo(E_0) \ino^2(E_0)}{\clc}}\,.
\end{equation}

\section{Decomposing Three Sources as Two}
\label{sec:threeastwo}

The discussions of decomposing two and three component backgrounds
included the tacit assumption that it was known how many source
classes were contributing to the background. In particular, a flat
plateau was always interpreted as a region where one source class
alone was contributing. However, such a flat baseline could still include
modulations within the error bars, and thus we wish to assess the
consequences of erroneously assuming a baseline is the signature of one dominant
source class rather than two.

Assume there exists a three-component spectrum such that components 1
and 2 are related by

\beq
\ \inb = \alpha \ina
~~~,
\eeq
where $\alpha$ is a possibly energy-dependent quantity. Then the total
fluctuation anisotropy can be written

\beq
\label{eq:3comp_alpha}
\ \ino^2 \clo = \left(\ino - \inc\right)^2 \frac{\cla + \alpha^2
  \clb}{\left(\alpha + 1\right)^2}+ \inc^2 \clc
~~~.
\eeq
If there is a region where component 3 is not contributing, and the
contributions from components 1 and 2 result in a plateau within
errors, then the fluctuation anisotropy $\clm$ for the sum of the
contributions inferred from the plateau is

\beq
\label{eq:clm}
\ \clm = \frac{\cla + \alpha^2 \clb}{\left(1 + \alpha\right)^2}
~~~,
\eeq
If $\alpha$ is energy-independent, then at all energies

\beq
\ \ino^2 \clo = \left(\ino - \inc\right)^2 \clm + \inc^2 \clc
~~~.
\eeq
This is analogous to Eq.~\ref{two}, where the contributions from
source classes 1 and 2 are treated as a single source class. Thus, the
intensity energy sprectrum of source class 3 can be determined exactly
using any of the above methods.

If $\alpha$ is energy-dependent, then the inferred value $\clm$ is not
valid at all energies and thus error will be introduced into the
decomposition. Since a statistically flat
plateau necessitates a small energy dependence on $\alpha$ over that
energy range, the resultant errors are small provided components 1 and
2 do not dramatically change spectral shape at energies beyond the
plateau. The magnitude of the error can be assessed via
Eq.~\ref{eq:3comp_alpha} for given assumptions on the spectral shapes of
the two components.

\section{Discussion}
\label{sec:disc}
Unraveling the contributions of multiple source classes to a diffuse background can be
accomplished in many cases by combining intensity and anisotropy
information. We have presented model-independent decomposition techniques which can recover the spectral shapes
of the constituents of a two-component diffuse background (low-anisotropy plateau,
known zero-anisotropy component) and techniques which can recover both
the shape and normalisation of the component spectra (double plateau,
high-anisotropy plateau, minimum, multiple $\ell$). Additionally, we
have discussed cases in which these techniques can be
applied to backgrounds of three or more components.

The techniques presented here are applicable at any wavelength and for
any diffuse background composed of uncorrelated source classes. For example, the recent Planck measurement of the anisotropy power spectrum of the cosmic infrared background (CIB) at multiple frequencies~\citep{Ade:2011ap} suggests an opportunity to apply the techniques of this paper.  Dusty, starforming galaxies are expected to be the primary contributor to the intensity and anisotropy of the CIB, and the Planck measurement of the CIB anisotropy has been used to constrain models of the starforming galaxy population.  The decomposition methods presented here offer an alternative means of investigating the composition of the CIB\@.

The ARCADE 2 experiment recently reported an excess in the temperature
of the cosmic radio background over the cosmic microwave background
temperature~\citep{Fixsen:2009xn}; the excess temperature is a factor
of $\sim5$ brighter than the expected contribution of radio point
sources~\citep{Fixsen:2009xn, Seiffert:2009xs}.  However, the
anisotropy of the excess emission is a factor of $\sim10$ smaller than
that of the CIB, which disfavors an interpretation of the excess as
emission from normal galaxies~\citep{Holder:2012nm}.  Dark matter
annihilation has been suggested as a possible origin of the radio
excess~\citep[e.g.,][]{Fornengo:2011cn}.  While there are currently
only upper limits on the anisotropy of the ARCADE excess, an eventual
detection of anisotropy would enable the decomposition techniques
presented here to be applied to understanding the origin of the radio
background. 

Because each method has a clear signature in the observed anisotropy energy spectrum and/or straightforward mathematical tests of validity, there is no ambiguity in selecting which technique to apply. With the exception of the known zero-anisotropy component technique, no {\em a priori} assumptions need to be made about the nature of the component spectra. Hence, the methods presented here are unique and model-independent.

Each technique detailed in this paper, however, requires the source classes to
have energy-independent fluctuation angular power, which is not a perfect
assumption for many
astrophysical sources, extragalactic ones in particular. Such a
dependence would introduce degeneracy into the
decompositions as any variations in the anisotropy energy spectrum
could be explained either by changes in the fractional makeup of the intensity,
as we have assumed here, or by changes in $\hat{C}_\ell$ with $E$ for
each source class. However, it is likely that these assumptions hold over at least some energy ranges. While in certain cases these methods can be applied to scenarios in which more than two components contribute to the diffuse emission, not all scenarios with three or more relevant contributors can be decomposed using the techniques presented here. Thus, a significant contribution from a third component in scenarios other than the two cases discussed in \S\ref{sec:threecomp} may render the application of these methods difficult at some energies.

The ability to decompose the intensity energy spectrum of the IGRB
and recover the constituent spectra, even at an accuracy of up to a multiplicative constant, 
is of paramount importance in
understanding the physical properties of the underlying source
classes. 
Specifically in the case
of blazars, the slope of their gamma-ray background contribution
reveals the spectral properties of the unresolved blazar
population~\citep{PavlidouVenters2007} or the relative contribution of
different type of blazars to the gamma-ray background and consequently to
the faint end of the blazar luminosity
function~\citep{Collaboration:2010gqa:srccounts, tonivaso11}.  If blazars are indeed a subdominant component of the gamma-ray background intensity, as suggested by recent constraints from the measured IGRB anisotropy~\citep{Cuoco:2012yf, Harding:2012gk}, such a decomposition would in principle allow us to deduce the slope of their collective emission at much higher accuracy than by modeling and subtracting the dominant components, the details of which may be largely unknown.
In addition, a decomposed intensity
spectrum for blazars can place constraints on the intensity of the EBL.  In the case of dark matter, a decomposed
intensity spectrum {\em is an uncontaminated measurement of the
 photon spectrum from dark matter annihilation or decay,} which in turn can provide information about the dark matter
particle mass and dominant annihilation or decay channels. 

At the same time, several of our techniques can constrain the source population anisotropy as well. Such constraints are also extremely important in understanding the statistical properties of a source class, as they provide information about the faint end of the luminosity function that is independent from that encoded by the collective intensity from unresolved members of the class. 

Taken in complement with other
analysis methods, these techniques can provide a unique, valuable
window through which to probe the physics of the IGRB or any
other diffuse astrophysical background. 

\section*{Acknowledgments}
BH acknowledges the Caltech Summer Undergraduate Research Fellowship (SURF)
  program, the Alain Porter Memorial SURF
  Fellowship, Barbara and Stanley Rawn, Jr., and the National Science Foundation
  Graduate Research Fellowship under Grant No.~DGE-0646086 for their generous
  support. BH also acknowledges CCAPP for its hospitality while this work was completed.  VP acknowledges support for this work provided by NASA through Einstein Postdoctoral Fellowship grant number PF8-90060 awarded by the Chandra X-ray Center, which is operated by the Smithsonian Astrophysical Observatory for NASA under contract NAS8-03060.  
JSG acknowledges support from NASA through Einstein Postdoctoral Fellowship grant PF1-120089 awarded by the Chandra X-ray Center, as well as from NSF CAREER Grant No.~PHY-0547102 (to John Beacom).  This work was partially supported by NASA through the Fermi GI Program grant number  NNX09AT74G. We are grateful to Shin'ichiro Ando and Tonia Venters for helpful discussions, and Anthony Readhead for valuable feedback.

\bibliography{decomp_v30}

\begin{thebibliography}{65}
\expandafter\ifx\csname natexlab\endcsname\relax\def\natexlab#1{#1}\fi

\bibitem[{Abazajian {et~al}\mbox{.}(2011)Abazajian, Blanchet, \&
  Harding}]{Abazajian:2010pc}
Abazajian K.~N., Blanchet S., Harding J., 2011, Phys.Rev., D84, 103007

\bibitem[{Abdo {et~al}\mbox{.}(2010{\natexlab{a}})Abdo
  {et~al.}}]{Collaboration:2010gqa:srccounts}
Abdo A., {et~al.}, 2010{\natexlab{a}}, Astrophys.J., 720, 435

\bibitem[{Abdo {et~al}\mbox{.}(2010{\natexlab{b}})Abdo
  {et~al.}}]{Abdo:2010nz:igrb}
Abdo A., {et~al.}, 2010{\natexlab{b}}, Phys.Rev.Lett., 104, 101101

\bibitem[{{Abdo} {et~al}\mbox{.}(2010){Abdo} {et~al.}}]{fermiebl}
{Abdo} A.~A., {et~al.}, 2010, Astrophys.J., 723, 1082

\bibitem[{{Ackermann} {et~al}\mbox{.}(2012){Ackermann}, {Ajello}, {Allafort},
  {Baldini}, {Ballet}, {Bastieri}, {Bechtol}, {Bellazzini}, {Berenji}, {Bloom},
  {Bonamente}, {Borgland}, {Bouvier}, {Bregeon}, {Brigida}, {Bruel}, {Buehler},
  {Buson}, {Caliandro}, {Cameron}, {Caraveo}, {Casandjian}, {Cecchi},
  {Charles}, {Chekhtman}, {Cheung}, {Chiang}, {Cillis}, {Ciprini}, {Claus},
  {Cohen-Tanugi}, {Conrad}, {Cutini}, {de Palma}, {Dermer}, {Digel}, {Silva},
  {Drell}, {Drlica-Wagner}, {Favuzzi}, {Fegan}, {Fortin}, {Fukazawa}, {Funk},
  {Fusco}, {Gargano}, {Gasparrini}, {Germani}, {Giglietto}, {Giordano},
  {Glanzman}, {Godfrey}, {Grenier}, {Guiriec}, {Gustafsson}, {Hadasch},
  {Hayashida}, {Hays}, {Hughes}, {J{\'o}hannesson}, {Johnson}, {Kamae},
  {Katagiri}, {Kataoka}, {Kn{\"o}dlseder}, {Kuss}, {Lande}, {Longo}, {Loparco},
  {Lott}, {Lovellette}, {Lubrano}, {Madejski}, {Martin}, {Mazziotta},
  {McEnery}, {Michelson}, {Mizuno}, {Monte}, {Monzani}, {Morselli},
  {Moskalenko}, {Murgia}, {Nishino}, {Norris}, {Nuss}, {Ohno}, {Ohsugi},
  {Okumura}, {Omodei}, {Orlando}, {Ozaki}, {Parent}, {Persic}, {Pesce-Rollins},
  {Petrosian}, {Pierbattista}, {Piron}, {Pivato}, {Porter}, {Rain{\`o}},
  {Rando}, {Razzano}, {Reimer}, {Reimer}, {Ritz}, {Roth}, {Sbarra}, {Sgr{\`o}},
  {Siskind}, {Spandre}, {Spinelli}, {Stawarz}, {Strong}, {Takahashi}, {Tanaka},
  {Thayer}, {Tibaldo}, {Tinivella}, {Torres}, {Tosti}, {Troja}, {Uchiyama},
  {Vandenbroucke}, {Vianello}, {Vitale}, {Waite}, {Wood}, \& {Yang}}]{SFlat}
{Ackermann} M. {et~al.}, 2012, \apj, 755, 164

\bibitem[{Ackermann {et~al}\mbox{.}(2011)Ackermann {et~al.}}]{Ackermann:2011wa}
Ackermann M., {et~al.}, 2011, Phys.Rev.Lett., 107, 241302

\bibitem[{Ackermann {et~al}\mbox{.}(2012{\natexlab{a}})Ackermann
  {et~al.}}]{Ackermann:2012uf}
Ackermann M., {et~al.}, 2012{\natexlab{a}}, Phys.Rev., D85, 083007

\bibitem[{Ackermann {et~al}\mbox{.}(2012{\natexlab{b}})Ackermann
  {et~al.}}]{Ackermann:2012rg}
Ackermann M., {et~al.}, 2012{\natexlab{b}}, arXiv

\bibitem[{Ade {et~al}\mbox{.}(2011)Ade {et~al.}}]{Ade:2011ap}
Ade P., {et~al.}, 2011, Astron.Astrophys., 536, A18

\bibitem[{Ando(2009)}]{Ando:2009fp}
Ando S., 2009, Phys.Rev., D80, 023520

\bibitem[{{Ando} \& {Komatsu}(2006)}]{ando_komatsu_06}
{Ando} S., {Komatsu} E., 2006, \prd, 73, 023521

\bibitem[{Ando {et~al}\mbox{.}(2007)Ando, Komatsu, Narumoto, \&
  Totani}]{Ando:2006cr}
Ando S., Komatsu E., Narumoto T., Totani T., 2007, Phys.Rev., D75, 063519

\bibitem[{Ando \& Pavlidou(2009)}]{Ando:2009nk}
Ando S., Pavlidou V., 2009, Mon.Not.Roy.Astron.Soc., 400, 2122

\bibitem[{Atwood {et~al}\mbox{.}(2009)Atwood {et~al.}}]{Atwood2009}
Atwood W.~B., {et~al.}, 2009, Astrophys.J., 697, 1071

\bibitem[{Bertone {et~al}\mbox{.}(2007)Bertone, Buchmuller, Covi, \&
  Ibarra}]{Bertone:2007aw}
Bertone G., Buchmuller W., Covi L., Ibarra A., 2007, JCAP, 0711, 003

\bibitem[{{Chakraborty} \& {Fields}(2012)}]{SFnachi}
{Chakraborty} N., {Fields} B.~D., 2012, ArXiv e-prints

\bibitem[{Cholis \& Salucci(2012)}]{Cholis:2012am}
Cholis I., Salucci P., 2012, Phys.Rev., D86, 023528

\bibitem[{Cuoco {et~al}\mbox{.}(2012)Cuoco, Komatsu, \&
  Siegal-Gaskins}]{Cuoco:2012yf}
Cuoco A., Komatsu E., Siegal-Gaskins J., 2012, arXiv

\bibitem[{Cuoco {et~al}\mbox{.}(2011)Cuoco, Sellerholm, Conrad, \&
  Hannestad}]{Cuoco:2010jb}
Cuoco A., Sellerholm A., Conrad J., Hannestad S., 2011,
  Mon.Not.Roy.Astron.Soc., 414, 2040

\bibitem[{{Dom{\'{\i}}nguez} {et~al}\mbox{.}(2011){Dom{\'{\i}}nguez},
  {Primack}, {Rosario}, {Prada}, {Gilmore}, {Faber}, {Koo}, {Somerville},
  {P{\'e}rez-Torres}, {P{\'e}rez-Gonz{\'a}lez}, {Huang}, {Davis},
  {Guhathakurta}, {Barmby}, {Conselice}, {Lozano}, {Newman}, \&
  {Cooper}}]{joelebl}
{Dom{\'{\i}}nguez} A. {et~al.}, 2011, Mon.Not.Roy.Astron.Soc., 410, 2556

\bibitem[{Faucher-Giguere \& Loeb(2010)}]{FaucherGiguere:2009df}
Faucher-Giguere C.-A., Loeb A., 2010, JCAP, 1001, 005

\bibitem[{{Fichtel} {et~al}\mbox{.}(1977){Fichtel}, {Hartman}, {Kniffen},
  {Thompson}, {Ogelman}, {Ozel}, \& {Tumer}}]{Fichtel77}
{Fichtel} C.~E., {Hartman} R.~C., {Kniffen} D.~A., {Thompson} D.~J., {Ogelman}
  H.~B., {Ozel} M.~E., {Tumer} T., 1977, \apjl, 217, L9

\bibitem[{Fields {et~al}\mbox{.}(2010)Fields, Pavlidou, \&
  Prodanovic}]{Fields:2010bw}
Fields B.~D., Pavlidou V., Prodanovic T., 2010, Astrophys.J., 722, L199

\bibitem[{{Finke} {et~al}\mbox{.}(2010){Finke}, {Razzaque}, \&
  {Dermer}}]{chuckebl}
{Finke} J.~D., {Razzaque} S., {Dermer} C.~D., 2010, Astrophys.J., 712, 238

\bibitem[{Fixsen {et~al}\mbox{.}(2009)Fixsen, Kogut, Levin, Limon, Lubin,
  {et~al.}}]{Fixsen:2009xn}
Fixsen D., Kogut A., Levin S., Limon M., Lubin P., {et~al.}, 2009

\bibitem[{Fornasa {et~al}\mbox{.}(2009)Fornasa, Pieri, Bertone, \&
  Branchini}]{Fornasa:2009qh}
Fornasa M., Pieri L., Bertone G., Branchini E., 2009, Phys.Rev., D80, 023518

\bibitem[{Fornasa {et~al}\mbox{.}(2012)Fornasa, Zavala, Sanchez-Conde,
  Siegal-Gaskins, Delahaye, {et~al.}}]{Fornasa:2012gu}
Fornasa M., Zavala J., Sanchez-Conde M.~A., Siegal-Gaskins J.~M., Delahaye T.,
  {et~al.}, 2012, arXiv

\bibitem[{Fornengo {et~al}\mbox{.}(2011)Fornengo, Lineros, Regis, \&
  Taoso}]{Fornengo:2011cn}
Fornengo N., Lineros R., Regis M., Taoso M., 2011, Phys.Rev.Lett., 107, 271302

\bibitem[{Fornengo {et~al}\mbox{.}(2004)Fornengo, Pieri, \&
  Scopel}]{Fornengo:2004kj}
Fornengo N., Pieri L., Scopel S., 2004, Phys.Rev., D70, 103529

\bibitem[{{Georganopoulos} {et~al}\mbox{.}(2010){Georganopoulos}, {Finke}, \&
  {Reyes}}]{markosebl}
{Georganopoulos} M., {Finke} J.~D., {Reyes} L.~C., 2010, \apjl, 714, L157

\bibitem[{Geringer-Sameth \& Koushiappas(2011)}]{GeringerSameth:2011iw}
Geringer-Sameth A., Koushiappas S.~M., 2011, Phys.Rev.Lett., 107, 241303

\bibitem[{Harding \& Abazajian(2012)}]{Harding:2012gk}
Harding J.~P., Abazajian K.~N., 2012, arXiv

\bibitem[{Holder(2012)}]{Holder:2012nm}
Holder G., 2012

\bibitem[{Hooper {et~al}\mbox{.}(2012)Hooper, Kelso, \&
  Queiroz}]{Hooper:2012sr}
Hooper D., Kelso C., Queiroz F.~S., 2012, arXiv

\bibitem[{Hooper \& Linden(2011)}]{Hooper:2011ti}
Hooper D., Linden T., 2011, Phys.Rev., D84, 123005

\bibitem[{Inoue(2011)}]{Inoue:2011bm}
Inoue Y., 2011, Astrophys.J., 733, 66

\bibitem[{{Inoue} \& {Totani}(2009)}]{inouetotani}
{Inoue} Y., {Totani} T., 2009, Astrophys.J., 702, 523

\bibitem[{Jungman {et~al}\mbox{.}(1996)Jungman, Kamionkowski, \&
  Griest}]{Jungman:1995df}
Jungman G., Kamionkowski M., Griest K., 1996, Phys.Rept., 267, 195

\bibitem[{{Kneiske} \& {Dole}(2010)}]{Kneiske+Dole_2010}
{Kneiske} T.~M., {Dole} H., 2010, \aap, 515, A19

\bibitem[{Knox(1995)}]{Knox:1995dq}
Knox L., 1995, Phys.Rev., D52, 4307

\bibitem[{Komatsu {et~al}\mbox{.}(2011)Komatsu {et~al.}}]{Komatsu:2010fb}
Komatsu E., {et~al.}, 2011, Astrophys.J.Suppl., 192, 18

\bibitem[{{Lacki} {et~al}\mbox{.}(2012){Lacki}, {Horiuchi}, \&
  {Beacom}}]{SFlacki}
{Lacki} B.~C., {Horiuchi} S., {Beacom} J.~F., 2012, ArXiv e-prints

\bibitem[{Mather {et~al}\mbox{.}(1990)Mather, Cheng, Shafer, Bennett, Boggess,
  {et~al.}}]{Mather+Cheng+Eplee+etal_1990}
Mather J.~C., Cheng E., Shafer R., Bennett C., Boggess N., {et~al.}, 1990,
  Astrophys.J., 354, L37

\bibitem[{Mazziotta {et~al}\mbox{.}(2012)Mazziotta, Loparco, de~Palma, \&
  Giglietto}]{Mazziotta:2012ux}
Mazziotta M., Loparco F., de~Palma F., Giglietto N., 2012, arXiv

\bibitem[{Miniati {et~al}\mbox{.}(2007)Miniati, Koushiappas, \&
  Di~Matteo}]{miniati_koushiappas_di-matteo_07}
Miniati F., Koushiappas S.~M., Di~Matteo T., 2007, Astrophys.J., 667, L1

\bibitem[{Overduin \& Wesson(2004)}]{Overduin:2004sz}
Overduin J.~M., Wesson P., 2004, Phys.Rept., 402, 267

\bibitem[{Pavlidou \& Venters(2008)}]{PavlidouVenters2007}
Pavlidou V., Venters T.~M., 2008, Astrophys.J., 673, 114

\bibitem[{{Primack} {et~al}\mbox{.}(2011){Primack}, {Dom{\'{\i}}nguez},
  {Gilmore}, \& {Somerville}}]{Primack+Dominguex+Gilmore+Somerville_2011}
{Primack} J.~R., {Dom{\'{\i}}nguez} A., {Gilmore} R.~C., {Somerville} R.~S.,
  2011, in American Institute of Physics Conference Series, Vol. 1381, American
  Institute of Physics Conference Series, {F.~A.~Aharonian, W.~Hofmann, \&
  F.~M.~Rieger}, ed., pp. 72--83

\bibitem[{Seiffert {et~al}\mbox{.}(2009)Seiffert, Fixsen, Kogut, Levin, Limon,
  {et~al.}}]{Seiffert:2009xs}
Seiffert M., Fixsen D., Kogut A., Levin S., Limon M., {et~al.}, 2009

\bibitem[{Siegal-Gaskins(2008)}]{SiegalGaskins:2008ge}
Siegal-Gaskins J.~M., 2008, JCAP, 0810, 040

\bibitem[{Siegal-Gaskins \& Pavlidou(2009)}]{Siegal-Gaskins2009}
Siegal-Gaskins J.~M., Pavlidou V., 2009, Phys.Rev.Lett., 102, 241301

\bibitem[{Siegal-Gaskins {et~al}\mbox{.}(2011)Siegal-Gaskins, Reesman,
  Pavlidou, Profumo, \& Walker}]{SiegalGaskins:2010mp}
Siegal-Gaskins J.~M., Reesman R., Pavlidou V., Profumo S., Walker T.~P., 2011,
  Mon.Not.Roy.Astron.Soc., 415, 1074S

\bibitem[{{So{\l}tan}(2007)}]{Soltan_2007}
{So{\l}tan} A.~M., 2007, \aap, 475, 837

\bibitem[{Spergel {et~al}\mbox{.}(2003)Spergel
  {et~al.}}]{Spergel+Verde+Peiris+etal_2003}
Spergel D., {et~al.}, 2003, Astrophys.J.Suppl., 148, 175

\bibitem[{{Sreekumar} {et~al}\mbox{.}(1998){Sreekumar}, {Bertsch}, {Dingus},
  {Esposito}, {Fichtel}, {Hartman}, {Hunter}, {Kanbach}, {Kniffen}, {Lin},
  {Mayer-Hasselwander}, {Michelson}, {von Montigny}, {Muecke}, {Mukherjee},
  {Nolan}, {Pohl}, {Reimer}, {Schneid}, {Stacy}, {Stecker}, {Thompson}, \&
  {Willis}}]{sreekumar_bertsch_dingus_etal_98}
{Sreekumar} P. {et~al.}, 1998, Astrophys.J., 494, 523

\bibitem[{Stecker {et~al}\mbox{.}(2007)Stecker, Baring, \&
  Summerlin}]{Stecker+Baring+Summerlin_2007}
Stecker F.~W., Baring M.~G., Summerlin E.~J., 2007, Astrophys.J., 667, L29

\bibitem[{Stecker \& Scully(2010)}]{floydebl}
Stecker F.~W., Scully S.~T., 2010, Astrophys.J., 709, L124

\bibitem[{Stecker \& Venters(2011)}]{Stecker:2010di}
Stecker F.~W., Venters T.~M., 2011, Astrophys.J., 736, 40

\bibitem[{{Stecker} \& {Venters}(2011)}]{SFtoni}
{Stecker} F.~W., {Venters} T.~M., 2011, \apj, 736, 40

\bibitem[{Steigman {et~al}\mbox{.}(2012)Steigman, Dasgupta, \&
  Beacom}]{Steigman:2012nb}
Steigman G., Dasgupta B., Beacom J.~F., 2012, Phys.Rev., D86, 023506

\bibitem[{{Strong} {et~al}\mbox{.}(2004){Strong}, {Moskalenko}, \&
  {Reimer}}]{Strong+Moskalenko_Reimer_2004}
{Strong} A.~W., {Moskalenko} I.~V., {Reimer} O., 2004, Astrophys.J., 613, 956

\bibitem[{Ullio {et~al}\mbox{.}(2002)Ullio, Bergstrom, Edsjo, \&
  Lacey}]{Ullio:2002pj}
Ullio P., Bergstrom L., Edsjo J., Lacey C.~G., 2002, Phys.Rev., D66, 123502

\bibitem[{{Venters} \& {Pavlidou}(2011)}]{tonivaso11}
{Venters} T.~M., {Pavlidou} V., 2011, Astrophys.J., 737, 80

\bibitem[{Venters \& Pavlidou(2012)}]{Venters:2012bx}
Venters T.~M., Pavlidou V., 2012, arXiv

\bibitem[{{Zhang} \& {Beacom}(2004)}]{zhang_beacom_04}
{Zhang} P., {Beacom} J.~F., 2004, Astrophys.J., 614, 37

\end{thebibliography}

\label{lastpage}

\end{document}